\documentclass{aa}  
\usepackage{amsmath}
\usepackage{graphicx}
\usepackage{graphics}
\usepackage{subfigure}
\usepackage{txfonts}
\usepackage{natbib}
\bibpunct{(}{)}{;}{a}{}{,}

\def\ltsima{$\buildrel<\over\sim$}
\def\lsim{\lower.5ex\hbox{\ltsima}~}
\def\gtsima{$\buildrel>\over\sim$}
\def\gsim{\lower.6ex\hbox{\gtsima}~}

\def\sfrlya{SFR$_{Ly\alpha}$}

\def\sfruv{SFR$_{UV}$}

\def\teff{\ifmmode T_{\rm eff} \else $T_{\mathrm{eff}}$\fi}

\def\lya{Ly$\alpha$} 
\def\ha{H$\alpha$} 
\def\hb{H$\beta$}

\def\ebv{$E_{B-V}$}

\def\fesc{$f_{esc}$}
\def\fescnorm{$f_{esc}^{norm}$}

\def\ergscm{erg~s$^{-1}$~cm$^{-2}$}
\def\cm2{cm$^{-2}$}

\def\ewlya{$EW_{\mathrm{Ly}\alpha}$}

\def\hi{H{\sc i}}
\def\hii{H{\sc ii}}
\def\nhi{$N$(H{\sc i})}
\def\oiii{O{\sc iii}}
\def\oii{O{\sc ii}}
\def\nii{N{\sc ii}}
\def\sii{S{\sc ii}}
\def\nh{\ifmmode N_{\mathrm{HI}}\else $N_{\mathrm{HI}}$\fi}

\def\izw{{\sc I}Zw 18}
\def\vexp{\ifmmode v_{\rm exp} \else v$_{\rm exp}$\fi}
\def\taua{\ifmmode \tau_{a}\else $\tau_{a}$\fi}

\begin{document}


%
   \title{On the influence of physical galaxy properties on \lya\ escape in star-forming galaxies}
  \subtitle{}
  \author{Hakim Atek \inst{1,2,3}
    \and Daniel Kunth \inst{2}
  \and Daniel Schaerer \inst{4,5}
  \and J. Miguel Mas-Hesse \inst{6}
  \and Matthew Hayes \inst{5,7}
  \and G\"oran \"Ostlin \inst{8}
  \and Jean-Paul Kneib \inst{1}
  }
  \institute{Laboratoire d'astrophysique, \'Ecole Polytechnique F\'ed\'erale de Lausanne, Observatoire de Sauverny, 1290 Versoix, Switzerland
  \email hakim.atek@epfl.ch
 \and Institut d'astrophysique de Paris, UMR7095 CNRS, Universit\'e Pierre \& Marie Curie, 98bis boulevard Arago, 75014 Paris, France
  \and Spitzer Science Center, California Institute of Technology, 1200 E. California Blvd, CA 91125, USA
  \and Observatoire de Gen\`eve, 51, Ch. des Maillettes, CH-1290, Sauverny, Switzerland
  \and CNRS, IRAP, 14, avenue Edouard Belin, F-31400 Toulouse, France
  \and Centro de Astrobiolog'a (CSIC-INTA), Departamento de Astrof'sica, POB 78, 28691 Villanueva de la Ca–ada, Spain
  \and Universit\'e de Toulouse, UPS-OMP, IRAP, Toulouse, France
   \and  Department of Astronomy, Oskar Klein Centre, Stockholm University, AlbaNova University Centre, SE-106 91 Stockholm, Sweden
   }
\authorrunning{Atek et al.}
\titlerunning{The regulation of \lya\ emission in star-forming galaxies}

\date{Received date; accepted date}

\abstract
{Among the different observational techniques used to select high-redshift galaxies, the hydrogen recombination line Lyman-alpha (\lya) is of particular interest as it gives access to the measurement of cosmological quantities such as the star formation rate of distant galaxy populations. However, the interpretation of this line and the calibration of such observables is still subject to serious uncertainties.}
{In this context, it important to understand the mechanisms responsible for the attenuation of \lya\ emission, and under what conditions the \lya\ emission line can be used as a reliable star formation diagnostic tool}
{We use a sample of 24 \lya\ emitters at $z \sim 0.3$ with an optical spectroscopic follow-up to calculate the \lya\ escape fraction and its dependency upon different physical properties. We also examine the reliability of \lya\ as a star formation rate indicator. We combine these observations with a compilation of \lya\ emitters selected at $z = 0 - 0.3$ from the literature to assemble a larger sample.}
{We confirm that the \lya\ escape fraction depends clearly on the dust extinction following the relation \fesc(\lya)$ = C_{Ly\alpha} \times 10^{-0.4~E(B-V)~k_{Ly\alpha}}$ where $k_{Ly\alpha} \sim 6.67$ and C$_{Ly\alpha} = 0.22$. However, the correlation does not follow the expected curve for a simple dust attenuation. A higher attenuation can be attributed to a scattering process, while \fesc(\lya) values that are clearly above the continuum extinction curve can be the result of various mechanisms that can lead to an enhancement of the \lya\ output. We also observe that the strength of \lya\ and the escape fraction appear unrelated to the galaxy metallicity. Regarding the reliability of \lya\ as a star formation rate (SFR) indicator, we show that the deviation of SFR(\lya) from the true SFR (as traced by the UV continuum) is a function of the observed SFR(UV), which can be seen as the decrease of \fesc(\lya) with increasing UV luminosity. Moreover, we observe redshift-dependence of this relationship revealing the underlying evolution of \fesc(\lya) with redshift.} 
{}   

  \maketitle

\section{Introduction}

The study of galaxy formation and evolution is strongly related to the observational techniques used to detect galaxy populations at different epochs. In particular, the detection of the \lya\ emission line in very distant galaxies has triggered many cosmological applications. The \lya\ line is predicted to be the dominant spectral signature in young galaxies \citep{charlot93,schaerer03}, and combined to its rest-frame wavelength in the UV, one obtains a very efficient tool to detect galaxies at $2 < z < 7$ in the optical window from ground-based telescopes. Therefore, this potential has often been used to assemble large samples of star-forming galaxies at high redshift \citep[e.g.][]{gronwall07,nilsson09,ouchi10,hayes10b,cassata11,kashikawa11}. \lya-based samples additionally offer the opportunity of constraining the stage of cosmic reionization using the expected sharp drop in their number density as the ionization state of the intergalactic medium (IGM) changes \citep[eg.][]{malhotra06}, or the shape of the \lya\ line profile itself \citep{kashikawa06,dijkstra10}.

The interpretation of the \lya-based observations hinges however on a complete understanding of the redshift-evolution of the intrinsic properties of galaxies, the physical properties responsible for \lya\ emission, and how the \lya\ emitters (LAEs) are related to other galaxy populations. Given the different selection methods, it is still unclear how the LAE population is connected to the Lyman-Break Galaxies (LBGs) selected upon their rest-frame UV continuum. The possible overlap between these galaxy populations has been questioned in many studies recently but no definitive answer has emerged yet. Specifically, only a fraction of LBGs \citep{shapley03} and \ha\ emitting galaxies \citep{hayes10b} show \lya\ in emission , and LBGs appear to be more massive, dustier and older \citep[e.g.][]{gawiser06, erb06}. A reasonable scenario attributes these observed differences to the radiative transport of \lya\ photons in the ISM \citep{schaerer08, verhamme08, pentericci10}. Most observed trends of \lya\ would be driven by variations of \nhi\ and the accompanying variation of the dust content that will dramatically enhance the suppression of the \lya\ line as shown in \citet{atek09a}. However, \citet{kornei10} found that \lya\ emission is stronger in older and less dusty galaxies, and \citet{finkelstein09c} found that \lya\ emission is stronger in massive and old galaxies. Although these results are still model-dependent, as they rely on SED fitting to infer the stellar population properties, they are apparently not compatible with the radiative transfer scenario.

We have been until recently in the curious situation, where we studied thousands of LAEs at redshift $z \sim 2$ and higher, whereas only a handful of \lya\ galaxies is available in the nearby universe without any statistical bearing, since such studies are hampered by the difficulty to obtain UV observations. In the past few years, the situation has changed significantly with the advent of the {\it Galaxy Evolution Explorer} \citep[GALEX, ][]{martin05} whose UV grism capabilities enabled the study of a large sample of UV-selected galaxies at $z\ \lsim 1$ \citep{deharveng08,atek09b,finkelstein09,scarlata09,cowie10}. The study of low-redshift LAEs offers the advantage of a much higher flux and the possibility to obtain critical complementary informations at longer wavelengths in subsequent optical follow-up observations. We obtained in \citet{atek09b} optical spectroscopy for a subsample of LAEs detected by {\it GALEX} at a redshift of $z \sim 0.3$. Using optical emission lines, we were able for the first time to empirically measure the \lya\ escape fraction \fesc(\lya) in a sample of LAEs and found a clear anti-corellation with the dust extinction. Other studies derived the same quantity at different redshifts and find in general the same trend \citep{scarlata09,kornei10,hayes10b}. Nevertheless, all these results show a large scatter around the \fesc(\lya)-dust relationship, which is presumably the result of a number of other physical parameters that may affect the \lya\ escape fraction.

The next generation of telescopes {\it James Webb Space Telescope} {\it JWST} and {\it Extremely Large Telescopes} ({\it ELTs}) will heavily depend on \lya\ as a detection and star-formation diagnostic tool in distant galaxies. Therefore, it is crucial that we understand how to retrieve the intrinsic \lya\ emission from the observed quantity in order to properly derive the cosmological quantities. Consequently, the \lya\ escape fraction represents a key-parameter for the interpretation of current and future \lya\ based surveys. Using our sample of $z \sim 0.3$ \lya\ galaxies, we discuss in the present work the dependence of the \lya\ escape fraction on the physical properties of the host galaxy. In Section \ref{sec:obs}, we present our spectroscopic follow-up of {\it GALEX}-detected LAEs and the sample of local galaxies included in the study. We expose in Section \ref{sec:prop} our emission-line measurements and the procedure we followed to derive the physical properties of our galaxies, i.e the metallicity and extinction estimates and the AGN/starburst classification. In Section \ref{sec:regulation}, we discuss the correlation of \lya\ with various parameters, while section \ref{sec:fesc} is devoted to the regulation of the \lya\ escape fraction. Finally we compare in section \ref{sec:sfr} the SFR measurement based on \lya\ with other indicators before a general summary in section \ref{sec:conclusion}.

\section{Observations }
\label{sec:obs}

\subsection{The GALEX sample}
\label{galex_sample_sec}
The galaxy sample used in the present study represents a subset of a \lya\ emitters sample detected by \citet{deharveng08} from a {\it GALEX} slitless spectroscopic survey. Details about grism mode and the spectral extraction are given in \citet{morrissey07}. The five deepest fields were used to extract all continuum spectra with a minimum signal-to-noise ratio (S/N) per resolution element of 2 in the far ultraviolet (FUV) channel  ($1350 \AA-1750 \AA$), giving a total area of 5.65 deg$^{2}$. \lya\ emitters are then visually selected on the basis of a potential \lya\ emission feature, which corresponds approximately to a threshold of \ewlya\ $\gsim 10 \AA$. This visual search yielded 96 \lya\ galaxies in the redshift range $z \sim 0.2 - 0.35$, classified into three categories ($1=$good, $2=$fair, and $3=$uncertain) according to the quality of their identification \citep[see Table 2. of ][]{deharveng08}.

\subsection{Spectroscopic follow-up}
\label{follow-up_sec}

We obtained optical spectra for 24 galaxies in the Chandra Deep Field South (CDFS) and ELAIS-S1, which have been first presented in \citet{atek09b}. We have selected only galaxies with good quality (Q=1 or 2) \lya\ spectra in \citet{deharveng08}. We used {\it EFOSC2} on the {\it NTT} at ESO La Silla, under good observational conditions, with photometric sky and sub-arcsec seeing ($0.5\arcsec - 1\arcsec$). We used Grism \#13 to obtain a full wavelength coverage in the optical domain ($3690-9320 \AA$) giving access to emission lines from [\oii] 3727\AA\ to [\sii] 6717+6731 \AA\ at the redshift range of our sample. A binning of $2 \times 2$ is used and corresponds to a pixel-scale of 0.24\arcsec\ px$^{-1}$ and a spectral resolution of FWHM $\sim$ 12 \AA\ (for 1\arcsec\ slit). To avoid second order contamination that affects the longer wavelength range, an order sorting filter has been mounted to cut off light blue-ward of 4200 \AA. We first used a 5\arcsec\ longslit to perform a spectrophotometry of our targets to obtain aperture matched fluxes with respect to the {\it GALEX} \lya\ measurements. Then, we re-observed our targets in spectroscopic mode with 1\arcsec\ slit, which offers the higher spectral resolution required to separate \ha\ from \nii\ contamination.

\begin{figure*}[htbp]
   \centering
   \includegraphics[width=19cm]{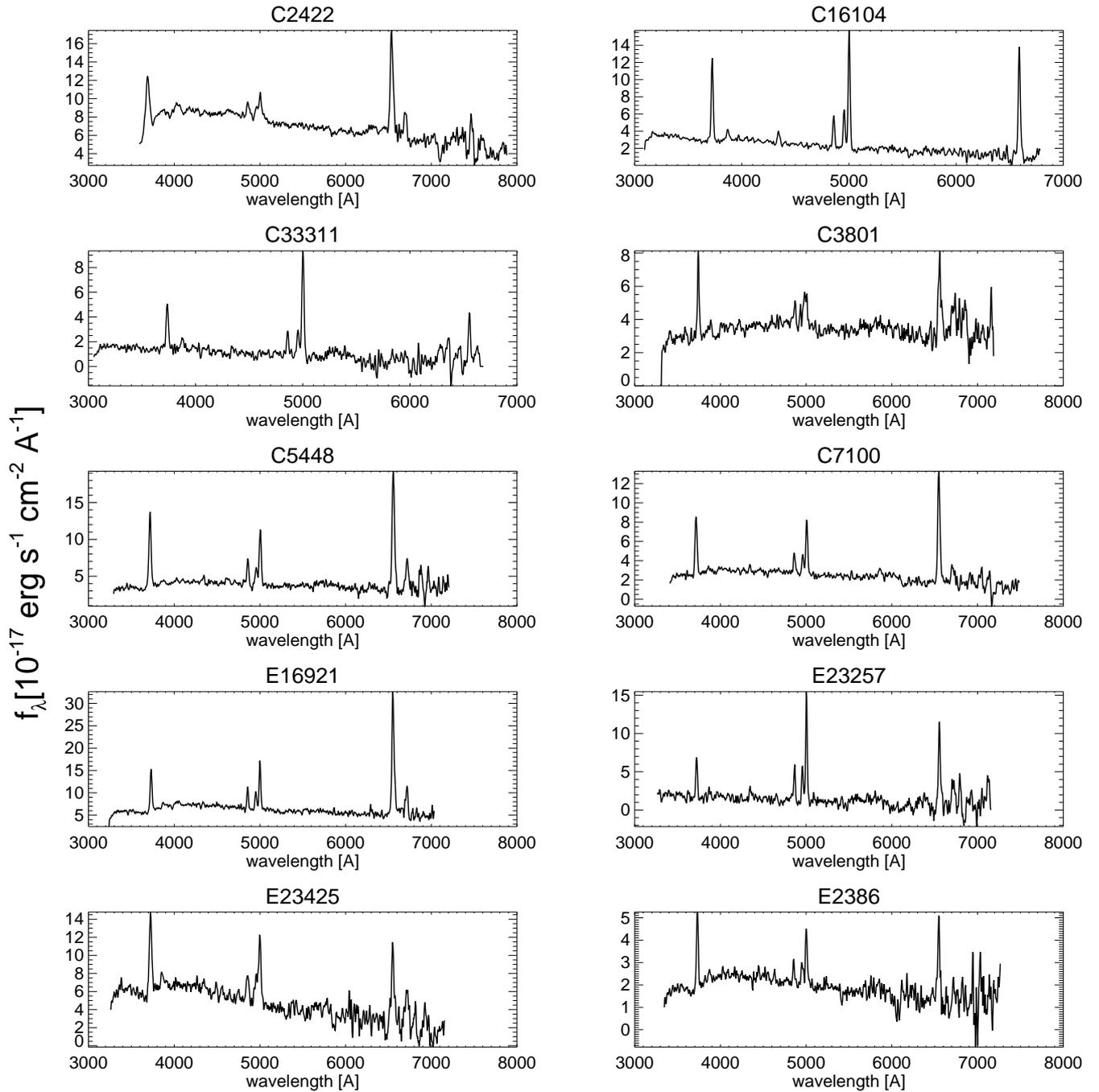}
   \vspace{-2cm}
   \caption{Examples of the rest-frame optical spectra of our sample illustrating the quality of the data. The flux density units are the same for all spectra, although the scales are different. The target name is marked on top of each spectrum.}
   \label{fig:galex_spectra}
\end{figure*}

The NTT spectra were reduced and calibrated using standard {\tt IRAF} routines. The EFOSC2 images were flat-fielded using both normal and internal flats taken before each pointing. The two flats give similar results, particularly in terms of fringes residual in the red part of the 2D spectra. Frames were combined for each object with cosmic ray rejection and useless segments of the images (blue-ward 4200 \AA\ for instance) removed to prevent artificial discrepancies in the sensitivity function and potential errors in the flux calibration. Then, the aperture extraction of 1D spectra was performed with the {\tt DOSLIT} task, where the dispersion solution is obtained from internal lamp calibration spectra to correct for telescope position variations. Finally, spectra were flux calibrated using a mean sensitivity function determined by observations of standard stars (Feige110, HILT600, LTT1020, EG21) from the \citet{oke90} catalog. A representative subset of the optical spectra is presented in Fig. \ref{fig:galex_spectra}.

\subsection{\it{IUE} starburst sample}

\begin{figure*}[htbp]
   \centering
   \includegraphics[width=19cm]{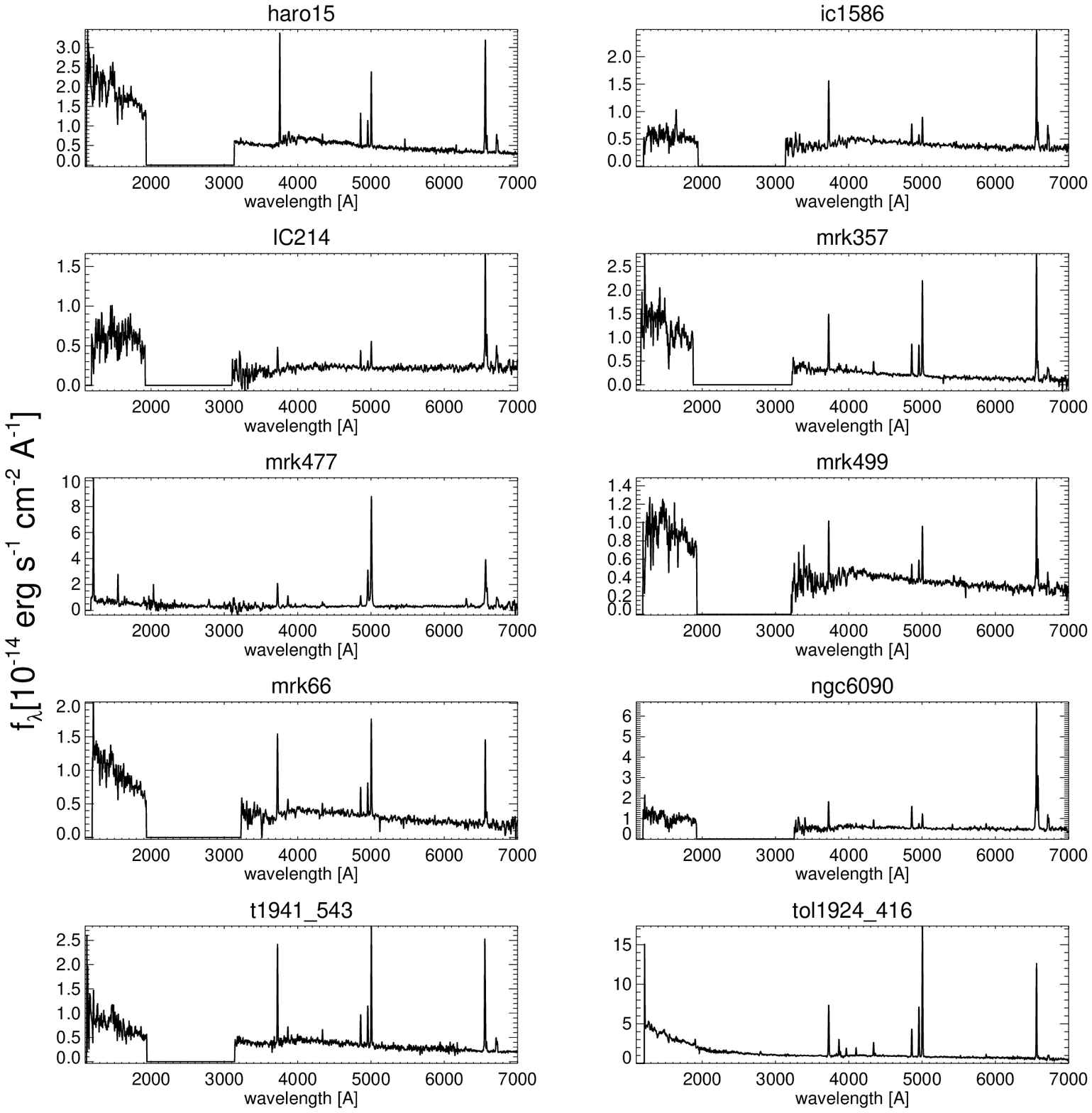}
      \vspace{-2cm}
   \caption{Example of rest-frame UV and optical spectra of the {\em IUE} sample. Most of the FUV spectra cover 1200-2000 \AA\ wavelength range and the optical spectra the 3200-7000 \AA\ range. The gap in between is set to zero.}
   \label{fig:iue_spectra}
\end{figure*}

We include in this analysis 11 local starbursts that have been spectroscopically observed in the UV by the {\it IUE} satellite. The UV spectra were obtained with 20\arcsec $\times$ 10\arcsec aperture, while the median size of the galaxies is about 25\arcsec\ in the optical). The spectra cover a wavelength range of $1200 - 3300$ \AA\ with a spectral resolution of 5 to 8 \AA. The ground-based optical spectra were obtained at {\it CTIO} ({\it Cerro Tololo Inter-American Observatory}) using 10\arcsec-wide slit and a spectral resolution of 8\AA. The spectra were extracted in a 20\arcsec-long aperture in order to match the {\it IUE} large aperture. We have re-analyzed the original UV spectra and the optical spectroscopic follow-up data presented in \citet{mcquade95} and \citet{storchi95} \citep[see also][]{giavalisco96}, and re-measured the line fluxes. The objects are chosen to be distant enough to separate the \lya\ feature of the galaxy from geocoronal \lya\ emission. The definition of a \lya\ emitter/absorber, as well as the \lya\ flux measurement, could be ambiguous for P Cygni profiles or an emission blended with absorption. Therefore, in our measurement procedure, we consider only the net \lya\ flux emerging from the galaxy. The line measurements have been performed following the same procedure used for our optical follow-up. We also show the 1D spectra in Fig. \ref{fig:iue_spectra}, which include UV and optical coverage. We included in our analysis only net \lya\ emitters, i.e. with \ewlya\ $> 0$. Furthermore, we included here our results from {\em HST}/ACS imaging of four nearby galaxies \citep{atek08}.

\begin{table*}[h]
\caption{Target informations and emission line measurements for the {\em GALEX} sample. Line fluxes are in units of 10$^{-15}$ \ergscm. The \lya\ flux is measured in the {\em GALEX} spectra by \citet{deharveng08}. The redshift is based on the mean position of optical lines with at least 3$\sigma$ significance. Target names follow \citet{deharveng08} nomenclature.} 
.\label{tab:measure}
\centering
\begin{tabular}{lccccccccl}
\hline\hline
Object    &  RA   & DEC    & z  & \lya & [\oii] & \hb   & [\oiii]$\lambda5007$ & \ha & [\nii]$\lambda6583$  \\
   &  (J2000)   & (J2000)    &  & &  &  &   &    &     \\
\hline   \\
CDFS--10937 & 53.7850 & -28.0454    &0.350 & 5.89 &  1.13$\pm$0.05    &  0.53$\pm$0.04   &   1.82$\pm$0.04  & 2.21$\pm$0.30  &...		  \\	 
CDFS--1348 & 53.2405 & -28.3883     &0.213 & 5.43 &  0.81$\pm$0.02    &  0.43$\pm$0.02   &   0.35$\pm$0.02  & 1.70$\pm$0.03  &0.53$\pm$0.02	  \\	   
CDFS--16104 & 53.2360 & -27.8879    &0.374 & 4.35 &  2.10$\pm$0.04    &  0.95$\pm$0.04   &   2.98$\pm$0.04  & 2.85$\pm$0.04  &0.24$\pm$0.03	 \\	  
CDFS--1821 & 53.2585 & -28.3577     &0.246 & 3.37 &  0.67$\pm$0.01    &  0.25$\pm$0.01   &   0.62$\pm$0.01  & 0.91$\pm$0.24  &0.13$\pm$0.03	  \\	   
CDFS--19355 & 53.7296 & -27.8008    &0.312 & 7.49 &  0.60$\pm$0.02    &  0.71$\pm$0.02   &   0.40$\pm$0.02  & 2.35$\pm$0.05  &1.12$\pm$0.04	  \\	  
CDFS--21667 & 53.2803 & -27.7424    &0.218 & 2.50 &   ...	      &  0.19$\pm$0.03   &    ...	    & 0.58$\pm$0.20  &0.27$\pm$0.02	  \\	   
CDFS--21739 & 53.7113 & -27.7293    &0.329 & 4.05 &  1.83$\pm$0.05    &  0.99$\pm$0.04   &   2.08$\pm$0.05  & 2.77$\pm$0.10  &0.72$\pm$0.10	 \\	  
CDFS--2422 & 52.8947 & -28.3395     &0.183 & 4.79 &  3.18$\pm$0.10    &  0.96$\pm$0.09   &   0.91$\pm$0.09  & 3.41$\pm$0.17  &1.19$\pm$0.08	  \\	   
CDFS--30899 & 53.3592 & -27.4543    &0.353 & 7.29 &   ...	      &  0.31$\pm$0.13   &    ...	    & 0.89$\pm$1.30  &...		  \\	  
CDFS--33311 & 53.1045 & -27.2904    &0.394 & 8.65 &  0.92$\pm$0.05    &  0.46$\pm$0.04   &   1.40$\pm$0.20  & 0.86$\pm$1.21  &0.02$\pm$0.09	\\	   
CDFS--3801 & 52.7375 & -28.2794     &0.288 & 3.62 &  1.05$\pm$0.05    &  0.73$\pm$0.06   &   0.87$\pm$0.10  & 1.48$\pm$0.30  &0.65$\pm$0.40	 \\	   
CDFS--4927 & 52.9765 & -28.2386     &0.282 & 8.12 &  0.24$\pm$0.01    &  0.21$\pm$0.01   &   0.10$\pm$0.01  & 0.47$\pm$1.00  &0.35$\pm$0.08	 \\	   
CDFS--5448 & 53.0780 & -28.2224     &0.286 & 11.4 &  3.25$\pm$0.07    &  1.23$\pm$0.06   &   2.25$\pm$0.08  & 5.14$\pm$0.30  &0.45$\pm$0.17	 \\	  
CDFS--6535 & 52.9622 & -28.1890     &0.214 & 5.16 &  ...	      &  0.16$\pm$0.04   &    ...	    & 0.46$\pm$0.35  &0.03$\pm$0.02	 \\	   
CDFS--6617 & 53.1743 & -28.1903     &0.212 & 17.2 &  2.41$\pm$0.05    &  1.42$\pm$0.04   &   7.61$\pm$0.04  & 3.85$\pm$0.09  &0.64$\pm$0.07	 \\	   
CDFS--7100 & 52.9993 & -28.1644     &0.245 & 2.96 &  1.81$\pm$0.05    &  0.73$\pm$0.05   &   1.56$\pm$0.05  & 3.33$\pm$0.16  &0.47$\pm$0.14	 \\	   
ELAISS1--16921 & 10.2733 & -43.8748 &0.317 & 4.14 &  2.64$\pm$0.07    &  1.50$\pm$0.06   &   2.61$\pm$0.06  & 5.49$\pm$0.11  &1.87$\pm$0.10	  \\  
ELAISS1--16998 &  9.5205 & -43.8745 &0.219 & 8.16 &   ...	      &  0.42$\pm$0.04   &	  ...	    & 1.10$\pm$0.03  &0.33$\pm$0.03	 \\   
ELAISS1--21062 &  9.6663 & -43.7225 &0.208 & 2.81 &  0.78$\pm$0.01    &  0.51$\pm$0.01   &   2.01$\pm$0.01  & 1.52$\pm$0.02  &0.10$\pm$0.01	 \\   
ELAISS1--23257 &  9.4752 & -43.6410 &0.296 & 5.17 &  1.29$\pm$0.06    &  1.13$\pm$0.06   &   3.12$\pm$0.06  & 3.36$\pm$0.26  &0.49$\pm$0.14	 \\   
ELAISS1--23425 &  9.3711 & -43.6356 &0.303 & 6.49 &  3.04$\pm$0.11    &  1.57$\pm$0.11   &   2.43$\pm$0.10  & 2.54$\pm$0.28  &0.64$\pm$0.16	 \\   
ELAISS1--2386 & 10.0078 & -44.4288  &0.275 & 2.81 &  0.87$\pm$0.05    &  0.33$\pm$0.04   &   0.67$\pm$0.04  & 0.90$\pm$1.20  &0.14$\pm$0.11	 \\   
ELAISS1--6587 &  9.5590 & -44.2436  &0.278 & 17.9 &  1.30$\pm$0.04    &  0.73$\pm$0.03   &   3.86$\pm$0.03  & 2.39$\pm$0.10  &0.12$\pm$0.05	  \\	
ELAISS1--8180 &  9.8839 & -44.1917  &0.186 & 10.1 &  0.46$\pm$0.04    &  0.47$\pm$0.10   &   0.15$\pm$0.03  & 1.58$\pm$0.80  &0.58$\pm$0.39	  \\

\hline
\end{tabular}                       
\end{table*}

\begin{table*}[h]
\caption{Same as Table 1 for the {\em IUE} sample. The line fluxes are in units of 10$^{-14}$ \ergscm. The \lya\ flux is measured in the {IUE} spectra.} 
\label{tab:iue_measure} 
\centering
\begin{tabular}{lccccccccl}
\hline\hline
Object    &  RA   & DEC    & z  & \lya & [\oii] & \hb   & [\oiii]$\lambda5007$ & \ha & [\nii]$\lambda6583$  \\
   &  (J2000)   & (J2000)    &  & &  &  &   &    &     \\
\hline   \\
Haro15	     &  00:48:35.4  &  -12:43:00    & 0.0210&  8.2  $\pm$ 0.4 & 25.91$\pm$0.16 &   7.88$\pm$0.17 &   18.72$\pm$0.17&  36.71$\pm$0.20&	5.13 $\pm$   0.21  \\	       
IC1586	     &  00:47:56.3  &  +22:22:22    & 0.0193&  2.5  $\pm$ 0.2 & 14.57$\pm$0.12 &   4.80$\pm$0.11 &    5.32$\pm$0.11&  21.39$\pm$0.09&	3.99 $\pm$   0.09   \\  	
IC214	     &  02:14:05.6  &  +05:10:24    & 0.0301& -9.1  $\pm$ 0.8 &  3.87$\pm$0.08 &   2.82$\pm$0.07 &    3.87$\pm$0.08&  15.43$\pm$0.06&	4.78 $\pm$   0.06   \\         
Mrk357	     &  01:22:40.6  &  +23:10:15    & 0.0528&  15.0 $\pm$ 1.0 & 14.00$\pm$0.06 &   7.85$\pm$0.06 &   21.63$\pm$0.07&  31.09$\pm$0.06&	4.72 $\pm$   0.06   \\  	
Mrk477	     &  14:40:38.1  &  +53:30:16    & 0.0381&  77.0 $\pm$ 0.9 & 23.75$\pm$0.20 &  12.32$\pm$0.22 &  121.00$\pm$0.21&  49.29$\pm$0.17&  20.77 $\pm$   0.16  \\	       
Mrk499	     &  14:40:38.1  &  +53:30:16    & 0.0256& -20.6 $\pm$ 0.8 &  6.68$\pm$0.07 &   2.42$\pm$0.07 &    6.24$\pm$0.07&  13.58$\pm$0.06&	3.49 $\pm$   0.06  \\		
Mrk66	     &  13:25:53.8  &  +57:15:16    & 0.0216&  7.1  $\pm$ 0.4 & 15.55$\pm$0.12 &   5.08$\pm$0.11 &   15.23$\pm$0.12&  13.53$\pm$0.09&	2.17 $\pm$   0.09   \\         
NGC5860      &  15:06:33.8  &  +42:38:29    & 0.0181&  4.7  $\pm$ 0.4 &  3.73$\pm$0.16 &   4.06$\pm$0.11 &    2.75$\pm$0.17&  31.47$\pm$0.09&  13.99 $\pm$   0.09 \\	
NGC6090      &  16:11:40.7  &  +52:27:24    & 0.0293&  11.0 $\pm$ 0.5 & 16.03$\pm$0.19 &  13.43$\pm$0.20 &    8.11$\pm$0.20&  65.86$\pm$0.15&  28.21 $\pm$   0.15  \\		
1941-543     &  19:45:00.5  &  -54:15:03    & 0.0193&  7.3  $\pm$ 0.9 & 20.73$\pm$0.21 &   7.74$\pm$0.22 &   25.31$\pm$0.21&  28.02$\pm$0.24&	2.53 $\pm$   0.25  \\		
Tol1924-416  &  19:27:58.2  &  -41:34:32    & 0.0098&  58.0 $\pm$ 0.4 & 71.97$\pm$0.17 &  36.73$\pm$0.15 &  173.50$\pm$0.16&  112.22$\pm$0.15&   2.50 $\pm$   0.14  \\             
\hline
\end{tabular}                       
\end{table*}

\section{Physical and spectral properties analysis}
\label{sec:prop}
\subsection{Emission line measurements}
\label{emission_sec}

The spectra were analyzed using the {\tt SPLOT} package in {\tt IRAF}. The redshifts of the galaxies were derived from the position of several emission lines, and line measurements performed interactively on rest-frame spectra. When possible, we measured the flux and equivalent width of [\oii] 3727\AA, [\oiii] 4959, 5007 \AA, H$\delta$, H$\gamma$, \ha, \hb\ and [\nii] 6548, 6584 \AA. For most spectra, the \ha\ line (6563 \AA) is blended with [\nii] lines (6548 and 6584 \AA) even for the 1\arcsec\ slit observations. In this case a deblending routine is used within {\tt SPLOT} measure individual fluxes in each line, then,  [\nii]/\ha\ line ratio is used to correct the spectrophotometric observations for [\nii] contamination. In cases where weak emission lines are only present in the higher resolution mode, photometry is deduced by scaling the flux with strong Balmer lines (such as \ha). Another concern is related to the contamination of Balmer emission lines by underlying stellar absorption. The equivalent width of stellar absorption is directly measured from higher order Balmer lines (typically H$\gamma$ and H$\delta$) provided they are in absorption. When these lines are in emission or undetectable, a typical average value of 2 \AA, representative of star forming galaxies, is adopted \citep{tresse96, gonzalez99}. 

To determine uncertainties in the line fluxes, we used {\tt SPLOT} package to run 1000 Monte Carlo simulations in which random Gaussian noise, based on the actual data noise, is added to a noise-free spectrum (our fitting model). Then, emission lines were fitted in each simulation. The computed MC errors depend essentially on the S/N quality of spectra. Note that these Monte Carlo simulations are done automatically in {\tt SPLOT after each line fit}.We then propagated the errors through the calculation of all the quantities described above and the line ratios, extinction etc, computed hereafter. 
  
\begin{figure*}[htbp]
   \centering
   \includegraphics[width=19cm]{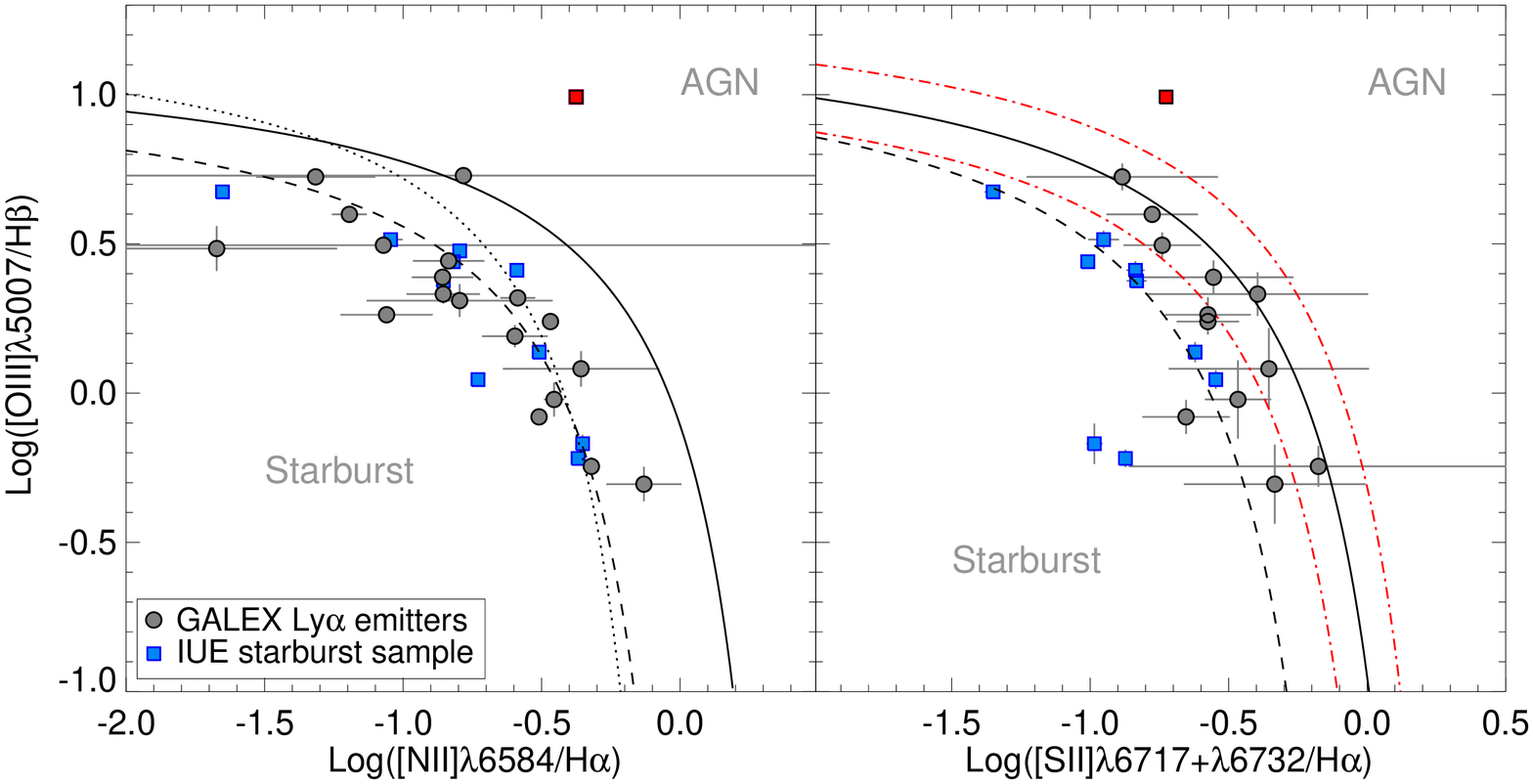}
   \caption{BPT diagrams used to classify narrow emission-line galaxies. GALEX \lya\ emitters are represented with black circles with corresponding error bars, together with an IUE sample of local starburst galaxies in blue squares. {\it The left panel} show  [\oiii] $\lambda5007$/\hb\ versus [\nii] $\lambda6584$/\ha\ ratios. The dashed and solid curves are theoretical boundaries separating starbursts and AGN objects assuming respectively an instantaneous burst \citep{dopita00} and an extended star formation episode \citep{kewley01}. The dotted curve is the ``Kauffmann line'' \citep{kauffmann03}. {\t The right panel} represents the [\oiii] $\lambda5007$/\hb\ versus [\sii] ($\lambda6717 + 6732$)/\ha\ diagnostic with the same theoretical models as in the right panel. The red dot-dashed shows typical model uncertainties of $\pm$ 0.1 dex. The diagram includes all the 11 {\it IUE} local galaxies, amongst which an object previously classified as an AGN (red square) that we decided to show as an example.}
   \label{fig:agn}
\end{figure*}

\subsection{Extinction}
\label{sec:extinction}

In order to measure the gas-phase dust extinction we use the Balmer ratio between \ha\ and \hb. The reddening contribution of our Galaxy is negligible since all our objects are located at high galactic latitude. The extinction coefficient $C(H\beta)$ is then given by the relation:
\begin{equation}
\label{c_equation}
\frac{f(H\alpha)}{f(H\beta)} = B \times 10^{- c [S(H\alpha) - S(H\beta)]} 
\end{equation} 
where f(\ha) and f(\hb) are the observed fluxes and $B$ is the intrinsic Balmer ratio. We adopted a value of $B=2.86$, assuming a case B recombination theory and a temperature of 10$^4$ K \citep{osterbrock89}. $S(H\alpha)$ and $S(H\beta)$ are the extinction curve values at \ha\ and \hb\ wavelengths respectively and were computed from the \citet{cardelli89} extinction law. The reddening \ebv\ is then simply computed using Eq. \ref{c_equation} and the relation $E(B-V) = c/1.47$. The extinction parameter A$_{V}$ is derived using a mean ratio of $R = A_{V} / E(B-V) = 3.2$.  

We will see in Sect. \ref{sec:dust} and \ref{sec:fesc} that some galaxies show negative E(B-V). Because we assumed an average correction for the stellar absorption, an overcorrection would lead to an underestimate of the \ha/\hb\ ratio that can explain the few negative values as it can be appreciated from the error bars on the E(B-V) values. In addition, a nebular reflection can also artificially increase the \hb\ contribution. In all the following calculations, we assigned E(B-V)=0 to all galaxies showing negative values. However, we show the actual measured values in the plots so the reader can appreciate the uncertainties.

\subsection{AGN--Starburst classification}
\label{sec:agn}

To discuss the nature of the ionizing source in our galaxies, we rely on optical emission line ratios. We examine the [\oiii] $\lambda5007$/\hb\ versus [\nii] $\lambda6584$/\ha\ diagnostic diagram, also known as the BPT diagram \citep{baldwin81}. It allows to distinguish \hii\ regions photoionized by young stars from regions which photoionization is dominated by a harder radiation field such as that of AGNs or low-ionization nuclear emission-line objects (LINERs). The underlying physics is that because photons from AGNs induce more heating than those of massive stars they will favor the emission from collisionally excited lines with respect to recombination lines. We also use an additional diagram originally proposed by \citet{veilleux87} representing  [\oiii] $\lambda5007$/\hb\ versus [\sii] ($\lambda6717 + 6732$)/\ha.   

Figure \ref{fig:agn} shows the location of our galaxies in these diagrams using dereddened line ratios, although these values are nearly insensitive to dust extinction, and associated uncertainties discussed above. A sample of IUE star forming galaxies in the nearby universe are also shown for comparison. As expected, the points lie in a relatively narrow region highlighting the separation of starburst-like galaxies from other types. In both figures, we have plotted the theoretical curves based on photoionization grids, that leave star forming regions to their lower-left and AGN-type objects to the upper-right. On the left panel, the dashed line represent the upper limit for \hii\ regions with an instantaneous zero-age star formation model \citep{dopita00}, the solid line for an extended burst scenario \citep[more than 4-5 Myr][]{kewley01}. The BPT boundary lines are strongly dependent on the effective temperature of the ionizing source, which evolves with time and therefore depends strongly on the star-formation time scale \citep[see for example][]{cervino94,kewley01}. Using a large sample from the sloan digital sky survey (SDSS), \citet{kauffmann03} revised this upper limit (plotted with a dotted line) downward. On the right panel we have also plotted the ``Dopita and Kewley lines'' for the second diagnostic method. However, these models are subject to different uncertainties related to the assumptions made on the chemical abundances, slope of the initial mass function (IMF) or the stellar atmospheres models. The errors on the starburst boundaries may be of the order of 0.1 dex \citep{kewley01}. We show in red dash-dotted line, an example of a typical error of 0.1 dex on the  [\oiii] $\lambda5007$/\hb\ versus [\sii] ($\lambda6717 + 6732$)/\ha.

We can see that all $z \sim 0.3$ galaxies but one lie on or below the solid starburst division line. This is very clear on the left  [\oiii]/\hb\ versus [\nii]/\ha\ diagram but proves more ambiguous on the right  [\oiii]/\hb\ versus [\sii]/\ha\ plot, although this is still consistent with both model and observations uncertainties. The observational uncertainties are particularly important on the right diagram because both redshifted [\sii] $\lambda6717, 6732$ and \ha\ lines are affected by a bright sky background at these wavelengths. Therefore, The BPT diagrams cast doubts on 3 candidates, as objects possibly excited by active nuclei, which we exclude from our analysis. This represents about $\sim 12\%$ of our sample, which is consistent with \citet{cowie10} and \citet{scarlata09} who find an AGN fraction of 10\% and 17\%, respectively, in their $z \sim 0.3$ LAE samples. This appears higher than typical AGN fraction ($\sim 0-5 \%$) found in high redshift LAE sample \citep{wang04,gawiser06,gawiser07,nilsson09,ouchi08}. Lacking optical spectroscopy, the high-redshift AGN fraction is determined from X-ray data, where the low-luminosity sources comparable to the low-z galaxies might be missed \citep{finkelstein09}. Therefore, this apparent evolution of the AGN fraction might be due to selection effects due to the difficulty to identify AGN in high redshift observations. At low redshift, \citet{finkelstein09} find a significantly higher fraction up to $\sim$ 45\% for their LAE sample. There is an overlap of 14 galaxies between the samples of \citet{cowie10} and \citet{finkelstein09}. Their AGN classification agree on nine of them, while the disagreement for the remaining three stems from, either a discrepancy in the line measurements, i.e. the position in the BPT diagram, or the presence/absence of high-ionization lines. In Particular, object GALEX1421+5239 has been classified as a star-forming galaxy by \citet{cowie11} based on the BPT diagram,  while \cite{finkelstein09} identified this object (EGS14 in their catalog) as an AGN after the detection of high-ionization lines.

\subsection{Oxygen abundance}  
\label{sec:metal}

The determination of the nebular metallicity can give important constraints on the chemical evolution and star formation histories of galaxies. The metal production and the chemical enrichment of the ISM is a direct result of stellar mass build-up and star formation feedback. In addition, the influence of the metal abundance on the evolution of the \lya\ output in these galaxies can be investigated.

In order to derive the nebular metallicity of our galaxies we use the optical emission line ratios and different metallicity calibrations according to the lines available in our spectra. For most of our galaxies, we measure the widely used abundance indicator, the $R_{23}$ index, which is based on the ratio ([\oii] $\lambda$3727 + [\oiii] $\lambda\lambda$ 4959,5007)/\hb\ \citep{pagel79, mcgaugh91, kobulnicky99, pilyugin03}. The main problem with the relationship (O/H)-$R_{23}$ is that there is a turnover around 12 + log(O/H) $\sim$ 8.4 that makes this index double-valued. A given $R_{23}$ value could correspond to a high or low metallicity. \citet{kewley08} have shown that additional line ratios such as [\nii]/[\oii] or [\nii]/\ha\ can be used to break this degeneracy. We used the [\nii]/[\oii] ratio to determine which of the upper and lower branch of the metallicity calibration to use.

Several studies have demonstrated that the $R_{23}$ method generally overestimates the actual metallicity when compared to the direct method \citep[e.g.][]{kennicutt03,bresolin05,yin07}. This discrepancy prevents us from having an accurate measurement with $R_{23}$ for intermediate metallicities, i.e. around the turnover region. Therefore, we here turn to other metallicity indicators that give more consistent results. We first use the emission-line ratio Log([\nii]$\lambda$6583/\ha), known as the $N2$ index \citep{denicolo02,pettini04}. The calibration of this indicator has been derived from the comparison of (O/H) measured using the ``direct'' method (via the electron temperature $T_{e}$), and the $N2$ index in a large sample of extragalactic \hii\ regions. They obtain the relationship that translates $N2$ to oxygen abundance, 12 + log (O/H) $= 8.90 + 0.57 \times N2$. Then, we consider the $O3N2$ index originally introduced by \citet{alloin79}, which corresponds to the line ratio Log{\{([\oiii] $\lambda$5007/\hb)/([\nii] $\lambda$ 6583/\ha)\}}. \citet{pettini04} used a sub-sample of their \hii\ regions to produce a calibration for this indicator and found a best fit with 12 + Log (O/H) $= 8.73 - 0.32 \times O3N2$. However this relationship is only valid for galaxies with $O3N2 < 2$, which is the cases for our galaxies. Of our sample of 21 galaxies, we were able to perform these metallicity measurements for 20 of them (for the remaining one the quality of the spectrum is not sufficient to confidently measure emission line fluxes), and for all of the 10 {\it IUE} galaxies. 

It is well known that the $R_{23}$ and other indicators are sensitive to the ionization parameter, and applying the above calibration to our sample assumes no variation in the ionization state, which can lead to inaccurate metallicity estimates. A detailed discussion about the effects of the ionization parameter on the different metallicity indicators has been presented in several studies using photoionization models \citep[see for example][]{cervino94, kewley02}. When $Z < 0.5 Z_{\odot}$, i.e. 12 + log (O/H) $\lesssim 8.6$, an increasing ionization parameter corresponds to an increasing value of the $R_{23}$ index, and inversely for metallicities higher than 0.5 $Z_{\odot}$ \citep[see Fig. 5 in][]{kewley02}. For single-valued indicators, the index increases with increasing ionization parameter. However, for their sample of $z \sim 0.3$ galaxies, \citet{cowie11} determined the metallicity using the ``direct'' method thanks to the detection of the [\oiii]$\lambda$4363, and compared the results to the $N2$ indicator. They found a good correlation with small scatter indicating relatively small effects from the ionization parameter. Moreover, while the absolute value of the metallicity may suffer from uncertainties, the trends analyzed here should hold irrespective of the absolute values, given the small range of metallicities covered by the sample.

\begin{figure*}[htbp]
   \centering
      \hspace{-0.3cm}
   \includegraphics[width=6.5cm]{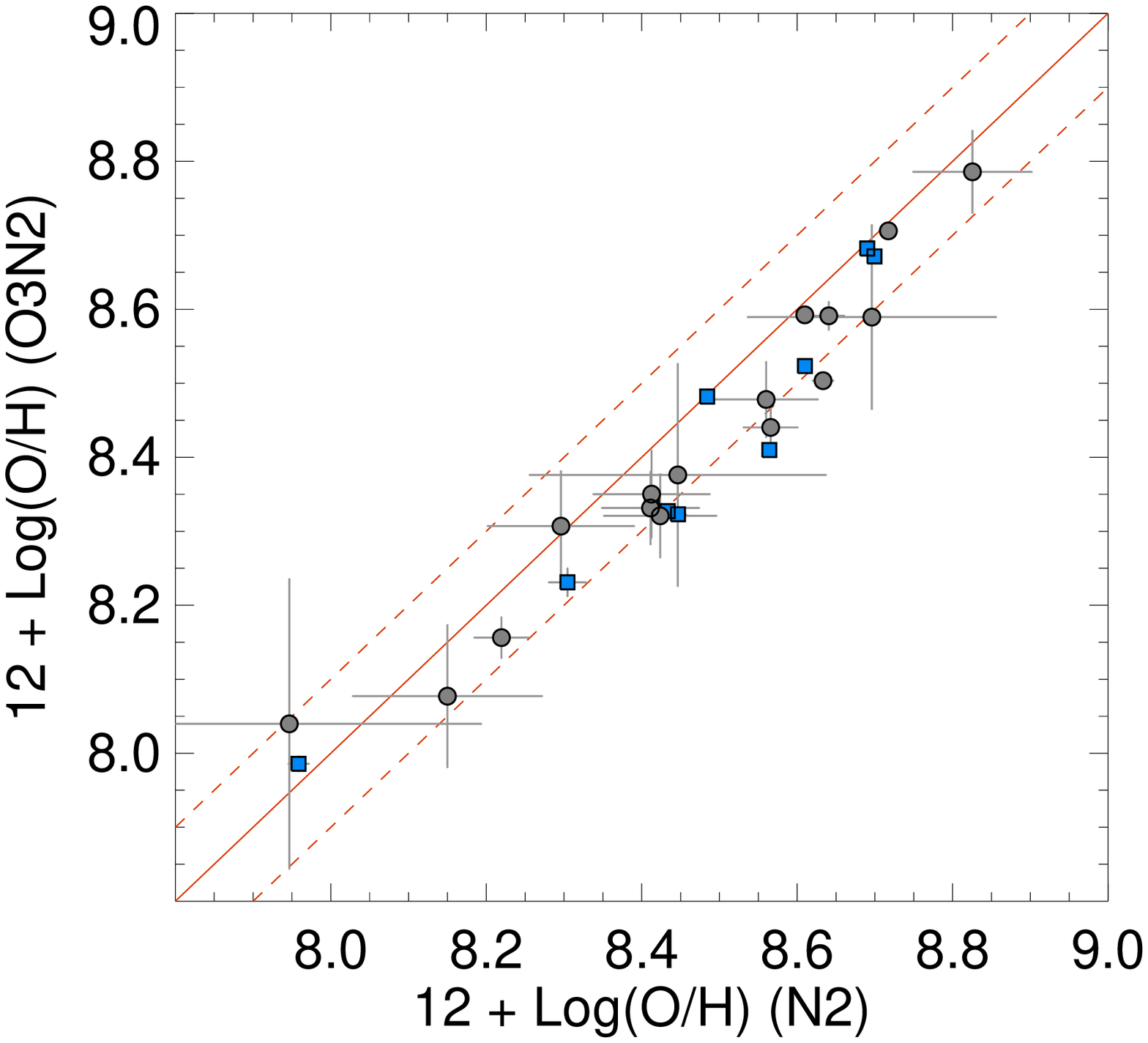}
   \hspace{-0.7cm}
    \includegraphics[width=6.5cm]{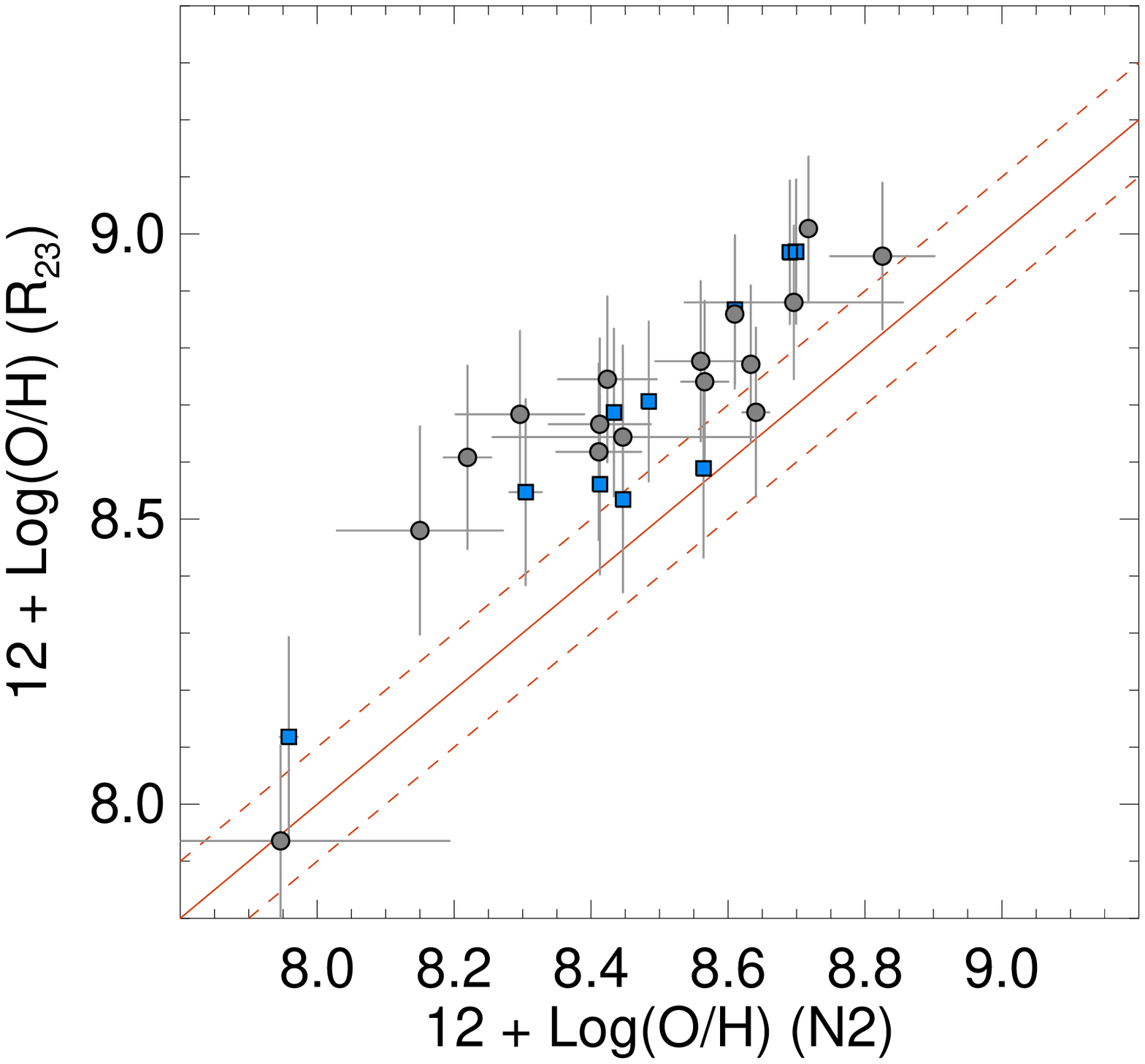}
    \hspace{-0.7cm}
     \includegraphics[width=6.5cm]{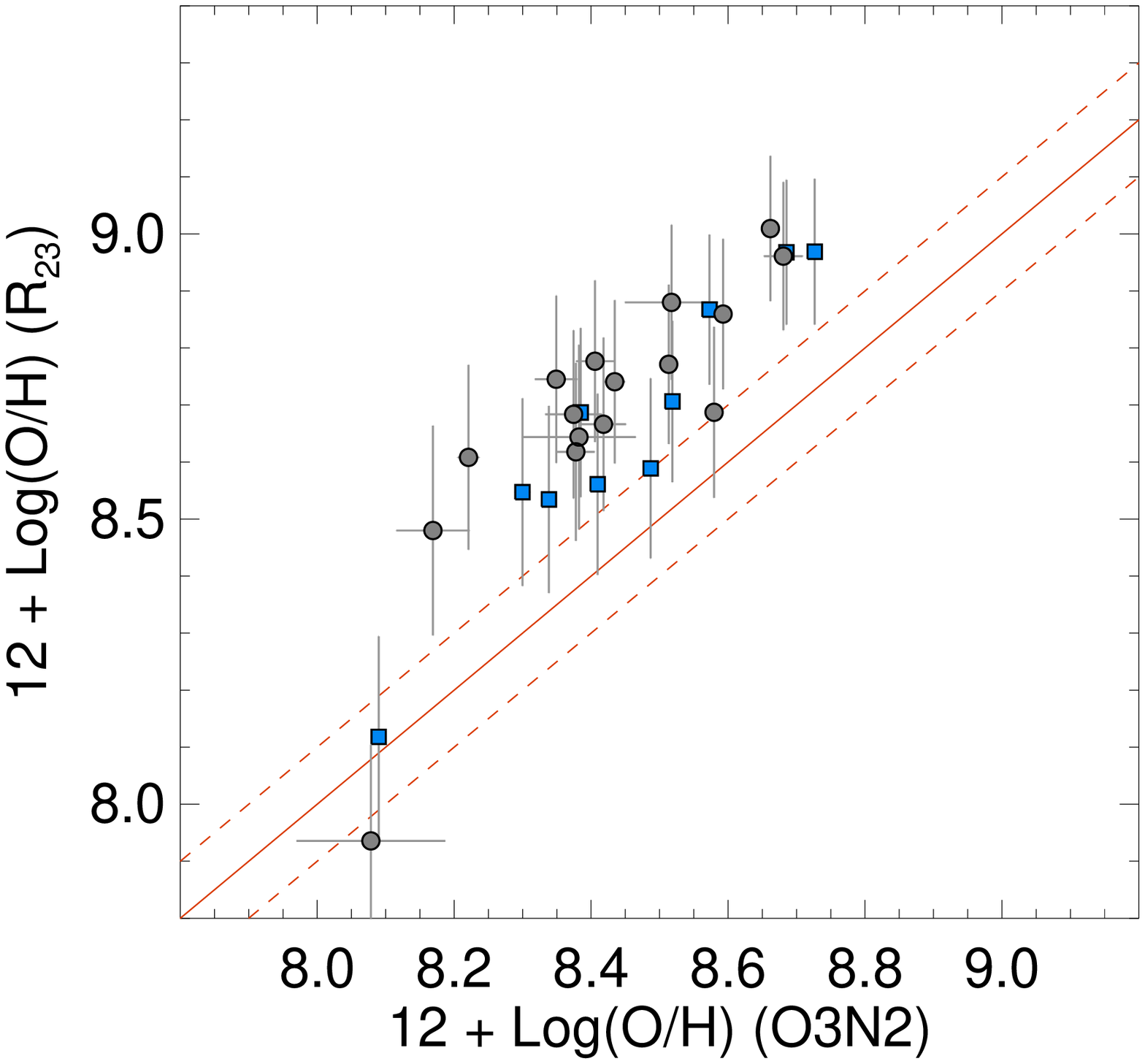}
        \hspace{-0.3cm}
   \caption{ \bf Comparison between the oxygen abundances derived different indicators: R$_{23}$= ([\oii] $\lambda$3727 + [\oiii] $\lambda\lambda$ 4959,5007)/\hb\ \citep{pagel79}, $N2$ = Log([\nii]$\lambda$6583/\ha) and $O3N2$ = Log{\{([\oiii] $\lambda$5007/\hb)/([\nii] $\lambda$ 6583/\ha)\}} \citep{pettini04}. The $z \sim 0.3$ galaxies are shown with black circles and the local galaxies with blue squares with 1$\sigma$ error bars. The solid line represents the 1:1 ratio with $\pm$0.1 dex uncertainties (dashed lines). Note that some galaxies are not present in all the plots because the corresponding diagnostic lines were not detected with a sufficient significance.}
   \label{fig:metal_comp}
\end{figure*}

We show in Fig. \ref{fig:metal_comp} a comparison between the oxygen abundance derived from several indicators: R$_{23}$,$N2$, and $O3N2$. It appears that $N2$ and $O3N2$ are in good agreement, although the values derived from $O3N2$ are systematically lower by 0.1 dex on average than those of the $N2$ method. On the other hand, although we do not have a direct measurement of the metallicity, we confirm that $R_{23}$ index overestimates the oxygen abundance by a factor up to 0.4 dex when compared to the two other methods used here. The $N2$ index is used in the rest of this analysis when it is available, otherwise $O3N2$ is adopted.

\section{The regulatory factors of the \lya\ output in galaxies}
\label{sec:regulation}

\subsection{Effects of metallicity}
\label{sec:ewlya_metal}

The first detections of the \lya\ emission line in star-forming galaxies goes back to the early {\em IUE} observations of nearby \hii\ galaxies \citep[e.g.][]{meier81, hartmann84, deharveng86}. However, in the case of detection, the line was very weak, while in the remaining galaxies it was found in absorprtion. The most natural explanation was the attenuation of the \lya\ photons by dust. The metallicity was often used as a dust indicator and compared with the \lya\ strength. The study of small galaxy samples \citep{meier81,terlevich93,charlot93} shows an anti-correlation between \ewlya\ and the metallicity, whereas the more comprehensive results of \citet{giavalisco96} show only a marginal trend.

\begin{figure}[htbp]
    \includegraphics[width=9cm]{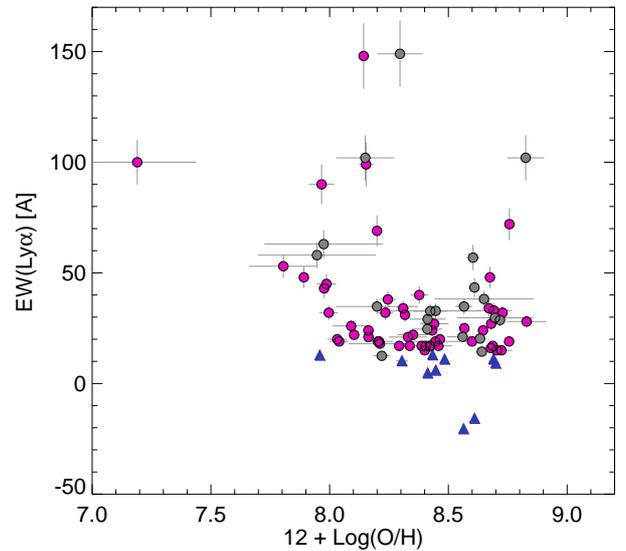}
   \caption{The \lya\ equivalent with as a function of metallicity. The oxygen abundance is derived from the $N2$ index Log([\nii]$\lambda$6583/\ha) using the equation of \citet{pettini04}. The present sample is shown with black circles, \citet{cowie11} galaxies with magenta circles, and IUE galaxies with blue triangles.}
   \label{fig:ewlya_metal}
\end{figure}

Here we investigate the effects of the metallicity on \lya\ emission in a large sample of galaxies. While the metallicity can be correlated with the \ha\ equivalent width as the result of metal enrichment with the age of the galaxy, this is not necessarily the case for the \lya\ strength. \citet{cowie11} compared the metallicity of a LAE sample and UV-selected galaxies and found that galaxies without \lya\ tend to have median metallicities 0.4 higher. \citet{Finkelstein11} also show that LAEs at low redshift have lower metallicities than similar-mass galaxies from the SDSS. One may expect a strong intrinsic \lya\ emission for a low-metallicity galaxy but the observed \lya\ output will also depend on other factors. Classical examples are \izw\ and SBS 0335, the most metal-poor galaxies known, that show a strong absorption in \lya\ \citep{kunth98,mashesse03,atek09a}. In fact, the metallicity distributions of the LAEs and UV-continuum galaxies measured in \citet{cowie11} overlap significantly. Figure \ref{fig:ewlya_metal} shows how \lya\ equivalent width is uncorrelated with the oxygen abundance. This means in particular that strong \lya\ emitters are not systematically metal-poor galaxies.

\subsection{Effects of dust}
\label{sec:dust}
As detailed in Sect. \ref{sec:extinction}, we have derived the dust extinction in the gas phase from the Balmer decrement \ha/\hb\ for all the galaxies studied here. Previous investigations concerning the effects of dust on the \lya\ emission, and in particular the correlation between the \lya\ equivalent width and extinction, have interestingly led to different results and conclusions. \citet{giavalisco96} found a large scatter in the correlation between \ewlya\ and the color excess $E(B-V)$ in a sample of 22 nearby starbursts. They concluded that, in addition to resonance scattering effects, \lya\  escape may be affected by the geometry of the neutral phase of the ISM, such as patchy dusty regions or ionized holes. However, these results stand in contrast with high redshift studies of \lya\ emitters. Composite spectra of $z \sim 3$ LBGs \citep{shapley03} showed a trend between \ewlya\ and $E(B-V)$ where objects with strong \lya\ emission have also steeper UV continuum. Similar trends have been observed more recently \citep{vanzella09,pentericci09,kornei10} where LBGs that exhibit \lya\ in emission tend to have bluer UV slopes than those with \lya\ in absorption. Following \citet{shapley03}, \citet{pentericci09} separated their $z \sim 4$ LBG sample in bins of extinction and noted same trend between the mean values of \ewlya\ versus stellar extinction. Again, all the observed correlations listed here are model-dependent, and may be affected by several uncertainties. Differences in the inferred physical properties may arise from the different extinction laws adopted in those studies, such as \citet{calzetti00} {\it vs} SMC \citep{prevot84}, which depends also on the age and the redshift of the galaxies \citep{reddy06, siana09}. More generally, the choice of a specific IMF and metallicity, or the degeneracy between age and extinction may introduce systematic uncertainties. Finally, as discussed in \citet{hayes13}, the {\em IUE} aperture at $z \sim 0.01$ probes a much smaller physical size than other studies at higher redshift. For lower dust extinction, \lya\ is spatially more extended leading to aperture losses, hence an underestimation of the \lya\ escape fraction.

\begin{figure}[htbp]
   \centering
   \includegraphics[width=9.5cm]{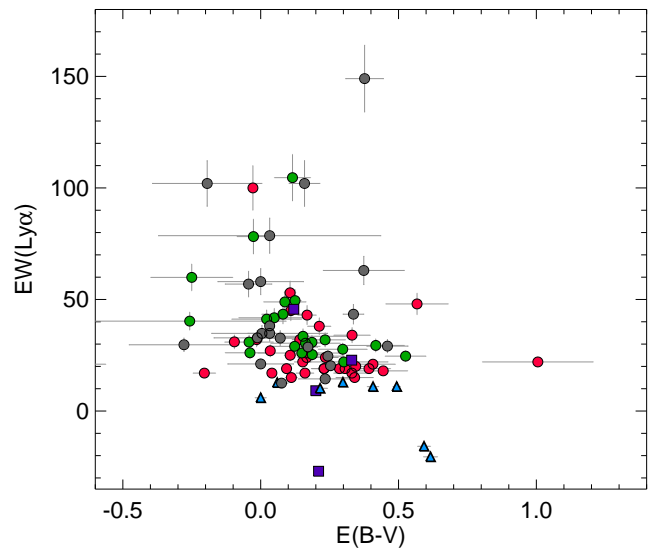}
   \caption{ The \lya\ equivalent width as a function of the gas-phase dust extinction. Black circles are the GALEX \lya\ emitters, green and red circles are the other GALEX samples of \citet{scarlata09} and \citet{cowie11}, respectively. The blue triangles represent the IUE spectroscopic sample, and purple squares local starbursts from \citet{atek08}.}
   \label{fig:ewlya_ebv}
\end{figure}

Figure \ref{fig:ewlya_ebv} shows \ewlya\ as a function of the extinction. The plot includes a compilation of local starbursts and $z \sim 0.3$ GALEX-selected samples. The published values of \citet{cowie11} are in bins of flux, which explains the discrete positions of E(B-V) for this sample. The \lya\ equivalent width shows a large scatter for $E(B-V)$ ranging from no extinction to $\sim 0.45$. This is consistent with previous results of \lya\ imaging observations of local starbursts \citep{atek08, ostlin09}, of which data points are plotted on the figure (purple squares). The dispersion of \lya\ strength observed for a given value of extinction is indicative of the influence of other galaxy parameters besides pure dust attenuation, such as \hi\ column density that may increase the suppression of \lya\ or neutral gas outflow that may ease the escape of \lya\ photons. However, as previously shown by \citet{schaerer08,verhamme08, atek09a}, the effects of superwinds might become insufficient in the case of high extinction, in which case we observe \lya\ in absorption or with a low equivalent width. This can be observed in Fig. \ref{fig:ewlya_ebv} where no high \ewlya\ values can be observed for large extinctions.

The absence of anti-correlation between EW$_{\rm{Ly}\alpha}$ and E(B-V) can be attributed to the difference in the type of extinction probed. While the stellar extinction is responsible for continuum attenuation, the \lya\ line flux is affected by the nebular extinction around the recombination regions. Therefore, the quantity of dust, and more importantly, the geometry of the dust distribution, can be very different between the two phases. In that case, the total extinction we derive from integrated fluxes would not be representative of the "effective" extinction affecting each radiation, and the total E(B-V) could become meaningless for very high extinction. We show in Table \ref{tab:extinction} the extinction factor at $\lambda$ = 1216 \AA\ for different E(B-V) values and two extinction laws \citep{cardelli89, prevot84}. As we can see, for E(B-V) values above 0.3, we could hardly detect any UV continuum nor \lya\ emission, since the extinction factor becomes rapidly higher than 100. At E(B-V) $\sim 1$, photons in the UV domain should be completely absorbed, which stands in contrast with \ewlya\ values around 50 \AA\ observed at such extinction levels. This suggests two mechanisms: (i) because of their resonant scattering, \lya\ photons can spatially diffuse far from the ionized regions where the Balmer lines are produced (from which we measure the extinction) and will sample different extinction, (ii) in the optical domain we will sample deeper regions and derive large average E(B-V) values, while in the UV, the observations are dominated by the contribution of the few stars/regions with lower values of extinction. Deeper regions would be completely obscured and would not contribute significantly in the UV. As a result, the average E(B-V) value in the UV (both \lya\ and continuum), would be much lower than in the optical \citep[see][for a detailed discussion about the differential extinction]{floranes12}. In Sect. \ref{sec:fesc}, we will also investigate the effects of the differential extinction scenario on the \lya\ escape fraction.

\begin{table}[htb]
\begin{center}
\centering
\caption{Extinction factors at $\lambda$ = 1216 \AA, calculated for different values of E(B-V) using the Galactic extinction \citep{cardelli89} and the SMC law \citep{prevot84}.}
\label{tab:extinction}
\renewcommand{\footnoterule}{}  
\begin{tabular}{l l c c c c}
\hline
\hline
\\
E(B-V)                           & 0.1  & 0.2 & 0.3 & 0.5 & 1 \\
\hline \\
\citet{cardelli89}  & 0.37   &     0.14  &    0.05   &  7$\times10^{-3}$  &     5$\times10^{-5}$                              \\
\citet{prevot84}       &  0.19    &    0.04   &   7$\times10^{-3}$  & 2$\times10^{-4}$     &    6$\times10^{-8}$               \\ 
\\
\hline  
\end{tabular}
\end{center}
\end{table}

\begin{figure}[htbp]
   \centering
   \includegraphics[width=9cm]{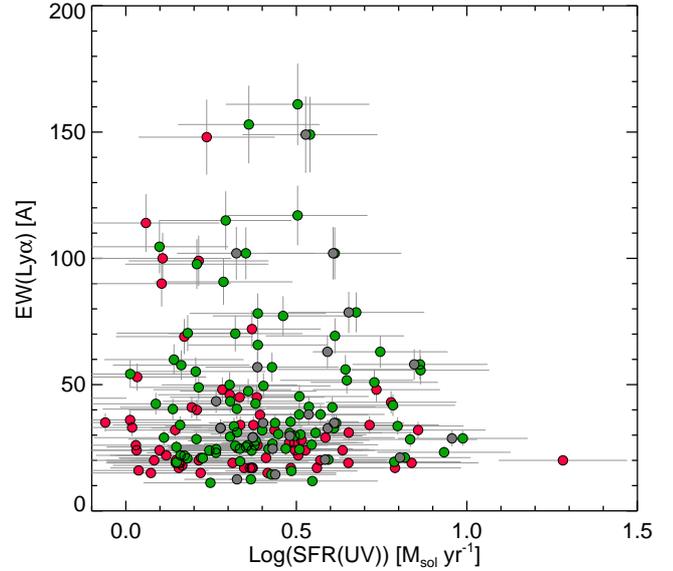}
   \caption{\ewlya\ as a function of UV star formation rate. The color-code is the same as Fig. \ref{fig:ewlya_ebv}}. The \lya\ equivalent width is the observed rest-frame value and the \sfruv\ values are calculated from FUV luminosity at 1530 \AA\ not corrected for extinction. \ewlya\ uncertainties are 10\% error bars, and \sfruv\ errors are derived from GALEX FUV photometry.
   \label{fig:ewlya_sfruv}
\end{figure}

\subsection{\lya\ equivalent width as a function of UV luminosity}

With the aim of understanding the different parameters governing the \lya\ escape and observability, many studies, both observational and theoretical, have found an apparent variation of the \lya\ strength with the UV luminosity \citep{shapley03,ando06,ouchi08,verhamme08,vanzella09,pentericci09,balestra10,stark10,schaerer11,shimizu11}. In particular, the data show a lack of high \lya\ equivalent width in luminous LBGs. A well known trend is the decrease of \ewlya\ with \sfruv\ \citep{ando04,tapken07,verhamme08}. At higher redshift, \citet{stark10} found that galaxies with strong \lya\ emitters have smaller UV luminosities, and have in general a bluer UV slope compared to galaxies with weak \lya\ emission. A similar plot is shown in Fig. \ref{fig:ewlya_sfruv}. Although, the correlation presented in \citet{verhamme08} or \citet{stark10} cover a larger range in both \sfruv\ and \ewlya, our data do not show such a clear trend, but rather an \ewlya\ distribution independent of UV luminosity. Together, Fig.\ \ref{fig:ewlya_ebv} and \ref{fig:ewlya_sfruv} indicate that the difference in the type of extinction probed combined with complex transport of \lya\ make the picture difficult to interpret. While the stellar extinction is responsible for continuum attenuation, the line flux is affected by the nebular extinction around the recombination regions combined to gas outflows. Therefore, the quantity of dust, and more importantly, the geometry of the dust distribution, can be very different between the two phases, and therefore between the continuum and the line behaviour. The absence of strong \lya\ emission at high SFR(UV) observed in high-$z$ samples can be due to differences in star formation histories and time scales, rather than an increase of the dust content, because powerful instantaneous bursts that produce very high \ewlya\ are more likely at young ages, i.e. small \sfruv, and constant star formation episodes at later stage characterized by modest \ewlya\ at equilibrium. 

\section{The \lya\ Escape Fraction}
\label{sec:fesc}

\begin{figure}[htbp]
   \includegraphics[width=9.1cm]{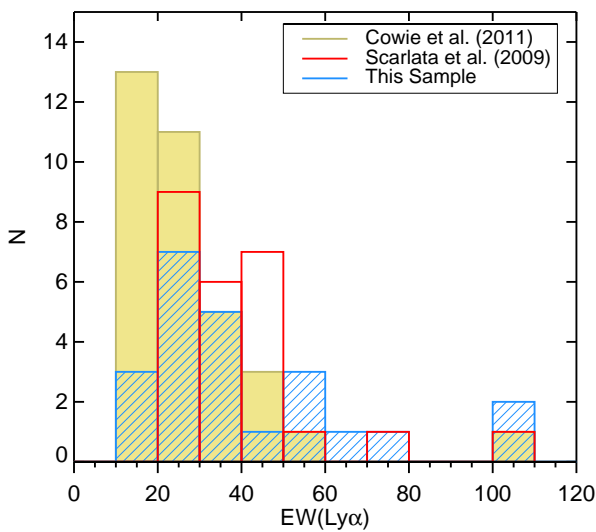}
   \caption{\lya\ equivalent width distribution for the three {\em GALEX} samples used in this work. It includes all sources with optical follow-up spectroscopy.}
   \label{fig:ewlya_histo}
\end{figure}

In \citet{atek09b}, we presented an empirical estimate of the \lya\ escape fraction in a large sample of galaxies selected from {\em GALEX} spectroscopy \citep{deharveng08}. Here we derive \fesc(\lya) for two additional $z \sim 0.3$ galaxy samples of \citet{scarlata09} and \citet{cowie11}. The \lya\ escape fraction is calculated following the equation: 
\begin{equation} 
 f_{esc}(\mathrm{Ly}\alpha) = f(\mathrm{Ly}\alpha)/(8.7 \times f(\mathrm{H}\alpha)_{C}) ,
  \label{eq:fesc}
\end{equation}       
where $f$(\lya) is the observed flux and $f$(\ha)$_{C}$ is the extinction-corrected \ha\ flux. The recombination theory (case B) gives a an intrinsic ratio of $f$(\lya)/$f$(\ha) = 8.7 \citep{brocklehurst71}. The Cowie et al. data were obtained with a 1\arcsec\ slit, while Scarlata et al. used a 1.5\arcsec slit. These measurements are not well matched to the {\em GALEX} aperture and can suffer from systematic errors. Both groups used a larger aperture for only few galaxies in their samples, and only to assess the importance of aperture effects. \citet{scarlata09} used a 5\arcsec slit for eight targets and compared the results with those made in the narrow slit, finding errors up to 25 \%. For the same reason, \citet{cowie11} compared their measurements in the narrow and wide slits with those of \citet{scarlata09} for a subset of galaxies that overlap, and they found a general agreement for their line fluxes. However in some cases the differences were important reaching a factor of 2 or more. We note that those large aperture measurements were not used in their analysis. If this is the case for some of our galaxies, we would overestimate their \fesc(\lya) values. Two effects can lead to inaccurate measurements of \fesc(\lya): (i) if the dust is not uniformly distributed across the galaxy, targeting only the central part of the galaxy would produce systematic error on E(B-V). (ii) The spatial extension of the \ha\ emission can go well beyond the size of the narrow slit, which results in an underestimate of the total \ha\ flux that should match the total \lya\ flux from {\em GALEX}. In order to better quantify these two effects on our results, we compared our measurements in 1\arcsec and 5\arcsec slits. We found that the dust reddening E(B-V) tends to be overestimated by a factor of 1.5 in the narrow slit, which possibly means the dust amount is higher in the center of the galaxy. Regarding \ha\ emission, we first corrected both observed \ha\ fluxes using the E(B-V) calculated in their respective slits. We found that the intrinsic \ha\ flux is underestimated by 30\% on average when using the narrow slit. We therefore applied these aperture correction factors to the samples of \citet{scarlata09} and \citet{cowie11}. The effect of such a correction is to increase the intrinsic \ha\ flux, and hence to decrease the \lya\ escape fraction. However, this does not have a significant impact on our results because Log(\fesc$_{{, \rm Ly}\alpha}$) is revised by only $\sim$0.1 on average, and the observed trends remain unchanged (eg. the correlation between Log(\fesc$_{{, \rm Ly}\alpha}$) and E(B-V)). Another difference is that, unlike the Scarlata et al. sample, Cowie et al. did not apply any stellar absorption to their emission lines. However, the median value for their \ha\ equivalent width is $\sim 80\AA$. A typical correction of 2\AA\ is then negligible and introduces a much smaller errors than the slit loss problems discussed above. Finally, we show in Fig. \ref{fig:ewlya_histo} a comparison between the \ewlya\ distribution of our sample and those of Cowie et al. and Scarlata et al. We note that the distribution of our sample is shifted to higher \ewlya\ compared to that of Cowie et al. which peaks at small \ewlya\ (less than 20 \AA). This is the result of our selection in the Deharveng et al. sample of good quality \lya\ lines only which favors high \ewlya. This will also result in selecting relatively higher \lya\ escape fraction as we will see later.

\begin{figure}[htbp]
   \includegraphics[width=9.1cm]{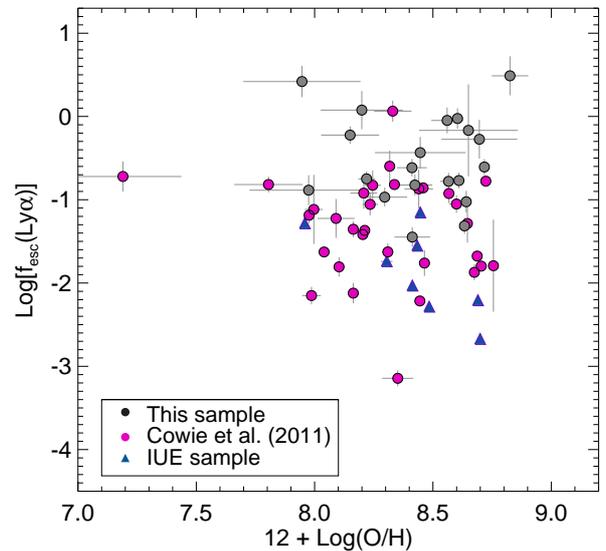}
   \caption{The \lya\ escape fraction as a function of the gas-phase metallicity. The oxygen abundance is derived from the $N2$ index Log([\nii]$\lambda$6583/\ha) using the equation of \citet{pettini04}. The color code as the same as in Fig. \ref{fig:ewlya_metal}}
   \label{fig:fesc_metal}
\end{figure}

We first show in Fig. \ref{fig:fesc_metal}, that the \lya\ escape fraction is not correlated to the oxygen abundance. This is similar to what we have seen in section \ref{sec:ewlya_metal}. It indicates that the metal abundance is not an important regulatory factor of the \lya\ escape.

\begin{figure*}[htbp]
   \centering
   \includegraphics[width=9.1cm]{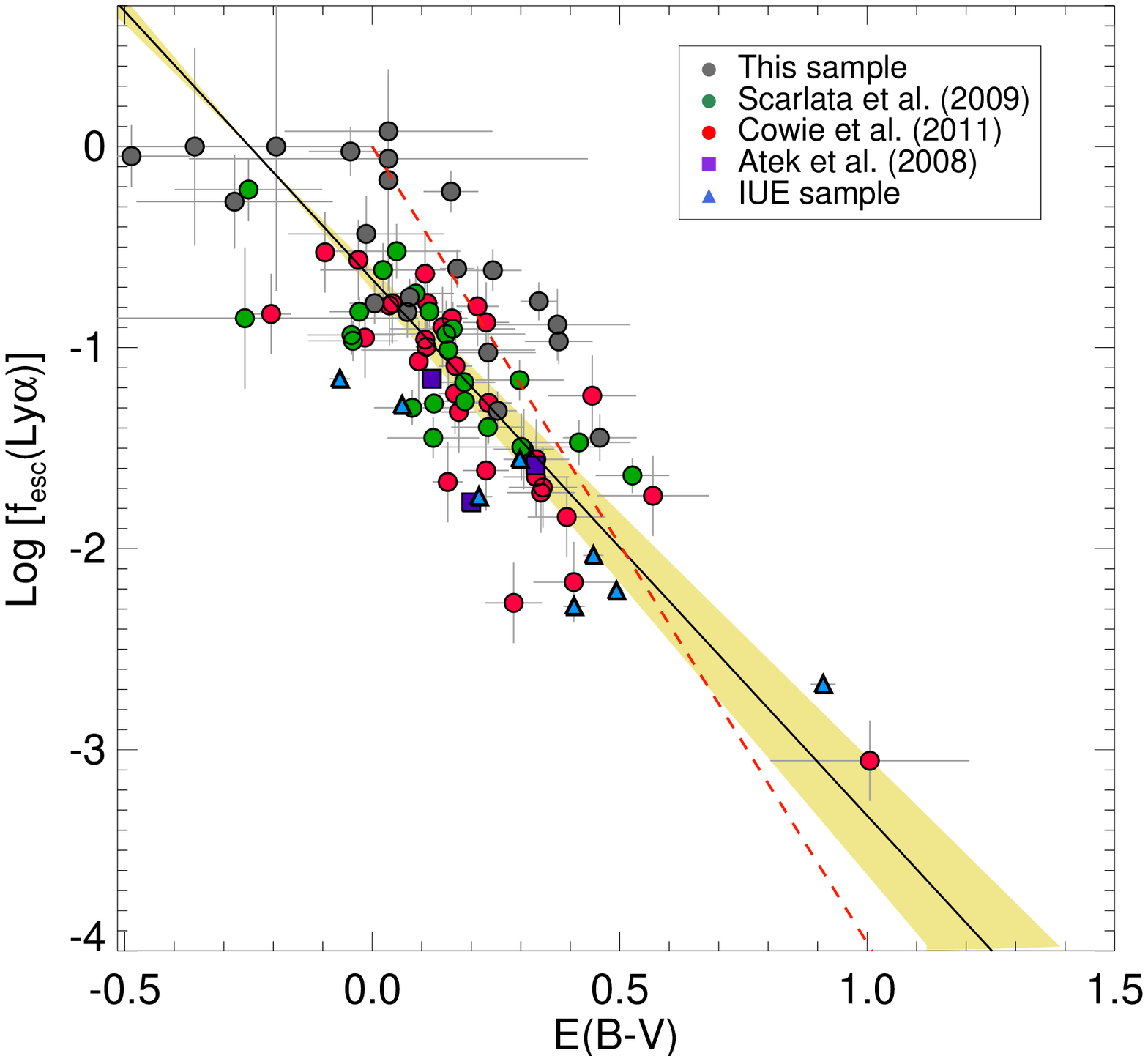}
    \includegraphics[width=9.1cm]{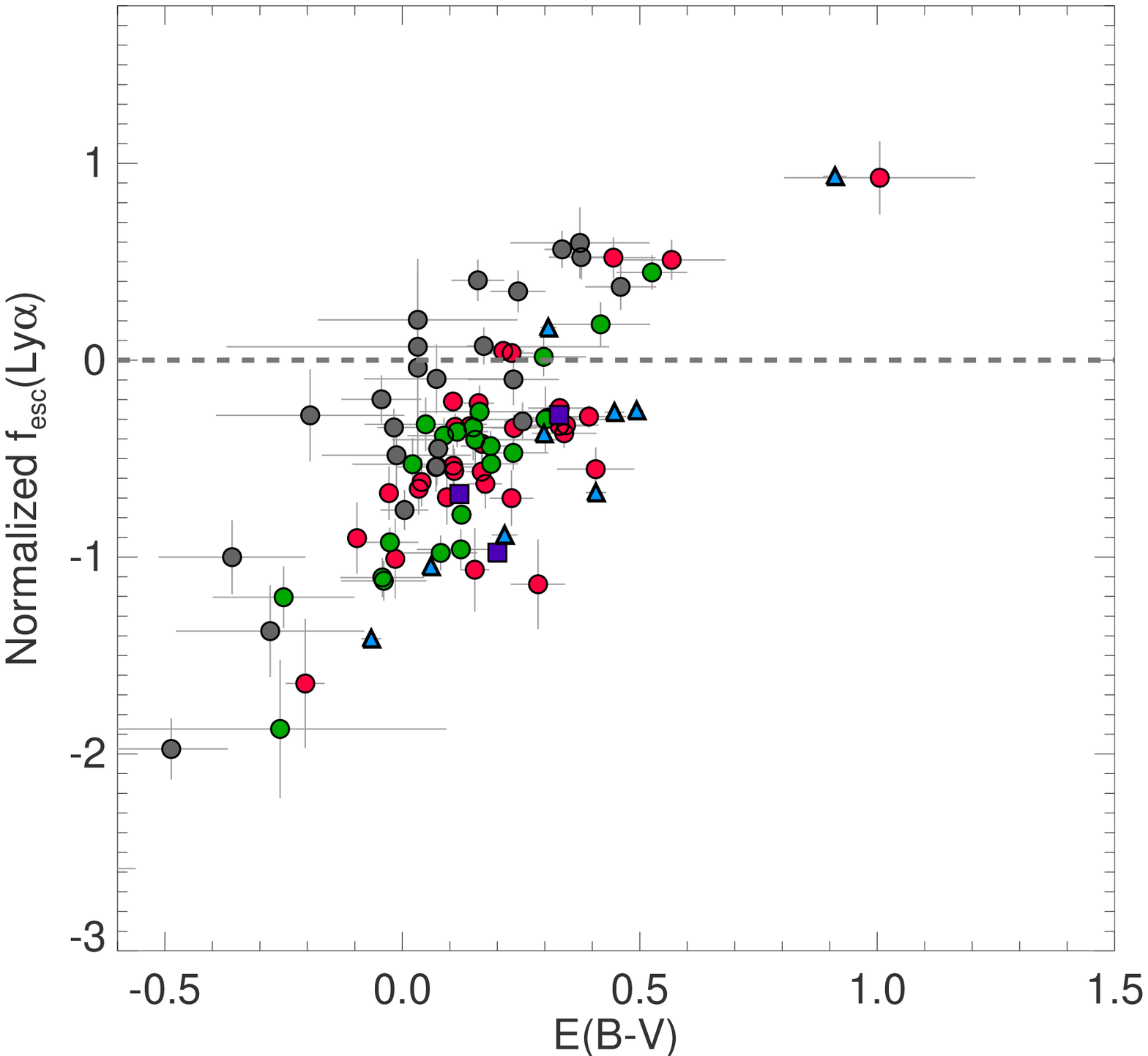}
   \caption{\lya\ escape fraction as a function of the nebular dust extinction. The figure shows the $z\sim 0.3$ \lya\ emitters of this work with black circles. We also derived \fesc(\lya) for other galaxy samples: $z \sim 0$ {\em IUE} sample represented by blue triangles, $z \sim 0.3$ sample of \citet{scarlata09} with green circles, \citet{cowie11} sample with red circles, $z \sim 0$ galaxies of \citet{atek08} with purple squares. Note that the negative values of E(B-V) are shown in the plot but were assigned E(B-V)=0 before \fesc(\lya) calculation or fitting the \fesc-E(B-V) relationship. The {\it left panel} shows \fesc(\lya) as a function of E(B-V) assuming a \citet{cardelli89} extinction law. The solid black line denotes the best 2-parameter fit to the relationship with both the slope and the intercept as free parameters. The yellow region covers the 1$\sigma$ uncertainties of the fit derived from MC simulations. The dashed red line is the expected attenuation law at the \lya\ wavelength. In the {\it right panel} we plot the normalized escape fraction \fesc$_{, rel}$(\lya)=log[\fesc(\lya)/\fesc(cont)] as a function of extinction. The normalized escape fraction represents the deviation of \fesc(\lya) from the classical dust attenuation law at 1216 \AA. The color code for the samples is the same as in the left panel.}
   \label{fig:fesc_ebv}
\end{figure*}

We present on the left side of Figure \ref{fig:fesc_ebv} the relation between \fesc(\lya) and the gas-phase extinction. The color-code for the different samples is presented in the legend. We fitted the Log(\fesc)-E(B-V) correlation using the {\tt MPFITEXY} IDL routine \citep{williams10} which is based on {\tt MPFIT} procedure \citep{markwardt09}. The routine fits the best straight line to the data taking into account the errors in both X and Y directions. The black solid line represents the best fit to the function  \fesc(\lya)$ = C_{Ly\alpha} \times 10^{-0.4~E(B-V)~k_{Ly\alpha}}$  where the origin C$_{Ly\alpha}$ and the slope $k_{Ly\alpha}$ are left as free parameters. Overall, the figure confirms the decreasing trend of \fesc(\lya) with increasing extinction observed in several studies \citep{atek08,atek09b,verhamme08,kornei10,hayes10b,hayes11}. However, when the origin is fixed at C$_{Ly\alpha} = 1$, i.e. assuming \fesc(\lya)=1 in the absence of dust, we derive an extinction coefficient of $k_{Ly\alpha} \sim 11.3 \pm 0.7$, which is slightly higher than what would be expected if dust extinction were the only factor affecting \lya\ escape, $k_{1216} = 9.9$ assuming a \citet{cardelli89} extinction law (the dot-dashed line).

 In the case of a 2-parameter fit, we find $k_{Ly\alpha} \sim 6.67 \pm 0.52$ and C$_{Ly\alpha} = 0.22 \pm 0.03$. The fit is shown by the solid black line in Fig. \ref{fig:fesc_ebv} while the uncertainties are represented by the shaded yellow region. To estimate the errors on the fit, we performed Monte Carlo simulations by generating a sample of 1000 datasets by randomly varying \fesc(\lya) and E(B-V) within their respective uncertainties. Then, we used the same procedure to fit each dataset. The two-parameter fit shows that even when no dust is present, the \lya\ emission is still attenuated by resonant scattering and the geometrical configuration of the ISM. Removing object GALEX1421+5239, which might be an AGN (cf. Sect. \ref{sec:agn}), results in a slightly higher extinction coefficient with $k_{Ly\alpha} \sim 6.73$. \citet{hayes11} used the same prescription to describe the effects of dust on the \lya\ escape fraction in a compilation of galaxies at redshift $z \sim 2$. However, the E(B-V) was derived from full SED fitting, and \fesc(\lya) was based either on the UV continuum or the \ha\ line corrected for extinction using the stellar E(B-V). After fitting the \fesc(\lya)-E(B-V) relation, they obtained $k_{Ly\alpha} \sim 13.8$ and C$_{Ly\alpha} = 0.445$, assuming a \citet{calzetti00} extinction law. Compared to these values, we obtain a smaller extinction coefficient $k_{Ly\alpha}$ and a lower intercept point at zero extinction. Here we draw attention to selection effects that can possibly contribute to the observed differences. The $z \sim 0.3$ LAEs detected by {\it GALEX} were selected based on a relatively strong \lya\ emission, which is therefore biased towards high \lya\ escape fractions. In the case of the $z \sim 2$ sample, galaxies were mostly selected upon their UV continuum or \ha\ emission. In fact, it has been shown in \citet{hayes10b} that \lya- selected galaxies have higher \fesc(\lya) values than \ha-selected ones.  
 
While the dust-dependence of \fesc(\lya) is clear, it does not follow classical dust attenuation prescriptions because of secondary parameters at play. In the right panel of Figure \ref{fig:fesc_ebv}, we illustrate the deviation of \fesc(\lya) from the value expected from E(B-V) under standard assumptions by plotting the normalized escape fraction defined as:
\begin{equation} 
 f_{esc}^{norm}(Ly\alpha) = f_{esc}(Ly\alpha) / f_{esc}(cont_{1216})
\end{equation}

where, \fesc(\lya) is the \lya\ escape fraction and $f_{esc}(cont_{1216})$ is the escape fraction of the UV continuum at 1216 \AA. The dashed horizontal line indicates the case where the \lya\ emission obeys to the same attenuation law as the UV continuum radiation. The values of \fesc(\lya) below this line can be explained by the the scattering of \lya\ photons in the neutral gas, which increases their optical path, thus the extinction coefficient. \fesc(\lya) values above the upper limit driven by the nebular dust extinction could be attributed to several physical processes or any combination thereof. First, the geometry of the ISM can change the behavior of \lya\ photons with respect to dust attenuation. In the case of a multi-phase ISM, where the neutral gas and dust reside in small clouds within an ionized inter-cloud medium, \lya\ photons scatter off of the surface of the clouds to escape easier than non-resonant photons that enter the clouds to encounter dust grains \citep{neufeld91, hansen06, finkelstein08}. Therefore, the preferential escape of \lya\ increases with increasing extinction, which is observed in our case. \citet{finkelstein11b} investigated the effects of ISM geometry on the \lya\ escape in a sample of 12 LAEs. Five of their galaxies where consistent with an enhancement of the observed \ewlya. \citet{laursen12} and \citet{duval13} recently discussed the scenario of multi-phase ISM scenario to explain unusually high \lya\ equivalent widths observed in high-$z$ LAEs. Their models show that very specific and restrictive conditions are needed in order to enhance the intrinsic EW(\lya), such as a high extinction and a low expansion velocity. Always in the context of a peculiar ISM geometry, \citet{scarlata09} discussed the possibility of a different effective attenuation resulting from a clumpy ISM \citep{natta84}, which does not take the classical e$^{- \tau}$ form but will depend on the number of clumps. This can reproduce the \lya\ enhancement observed in the objects above the dashed line. Given the increase of \fescnorm\ with increasing extinction, it is possible that the two mechanisms (enhancement of \lya\ by scattering and a peculiar extinction law) are at play, as they both require a high extinction to significantly increase the escape of \lya. If we go beyond the simplistic spherical geometry, more realistic configurations of the ISM could also favor \lya\ transmission. Ionized cones along the observer's sightline can preferentially transmit \lya\ photons via channeling, i.e. scattering of \lya\ photons on the \hi\ outskirts along the cone. Moreover, the \lya\ strength depends also on the inclination of the galaxy. Because \lya\ photons will follow the path of least opacity they will preferentially escape face-on. This is not necessarily the case for \ha\ photons which don't undergo an important angular redistribution \citep{verhamme12}. 

More generally, as we have shown in Sect. \ref{sec:dust}, when the extinction is high enough (typically E(B-V) $> 0.3$), the differential extinction sampling would naturally produce \fescnorm\ values higher than one. In the \hii\ regions affected by dust, \fesc(\lya) will be very low, while in the dust-free regions, where the \lya\ emission will dominate the rest of Balmer lines, \fescnorm\ will be higher than unity, since the total E(B-V) derived does not apply to those regions. In summary, if one expect starburst galaxies to have a gradation of extinction in different parts of the \hii\ regions, from regions completely onscured in the UV to regions almost devoid of dust, the convolution of all regions would give an evolution of \fescnorm\ with average E(B-V) similar to the diagram shown in Fig. \ref{fig:fesc_ebv}.

\begin{figure*}[htbp]
   \centering
     \includegraphics[width=9cm]{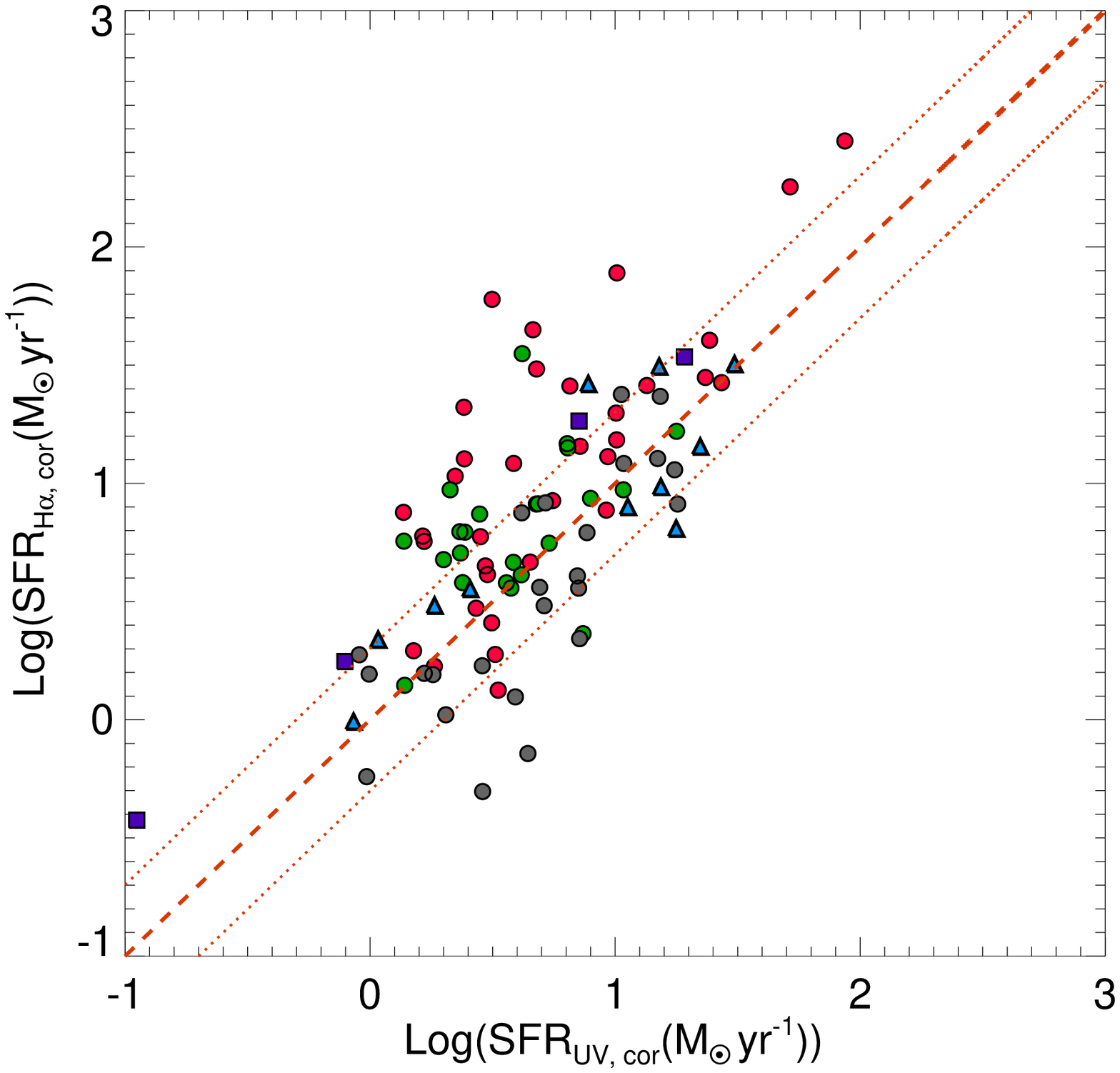}
   \includegraphics[width=9cm]{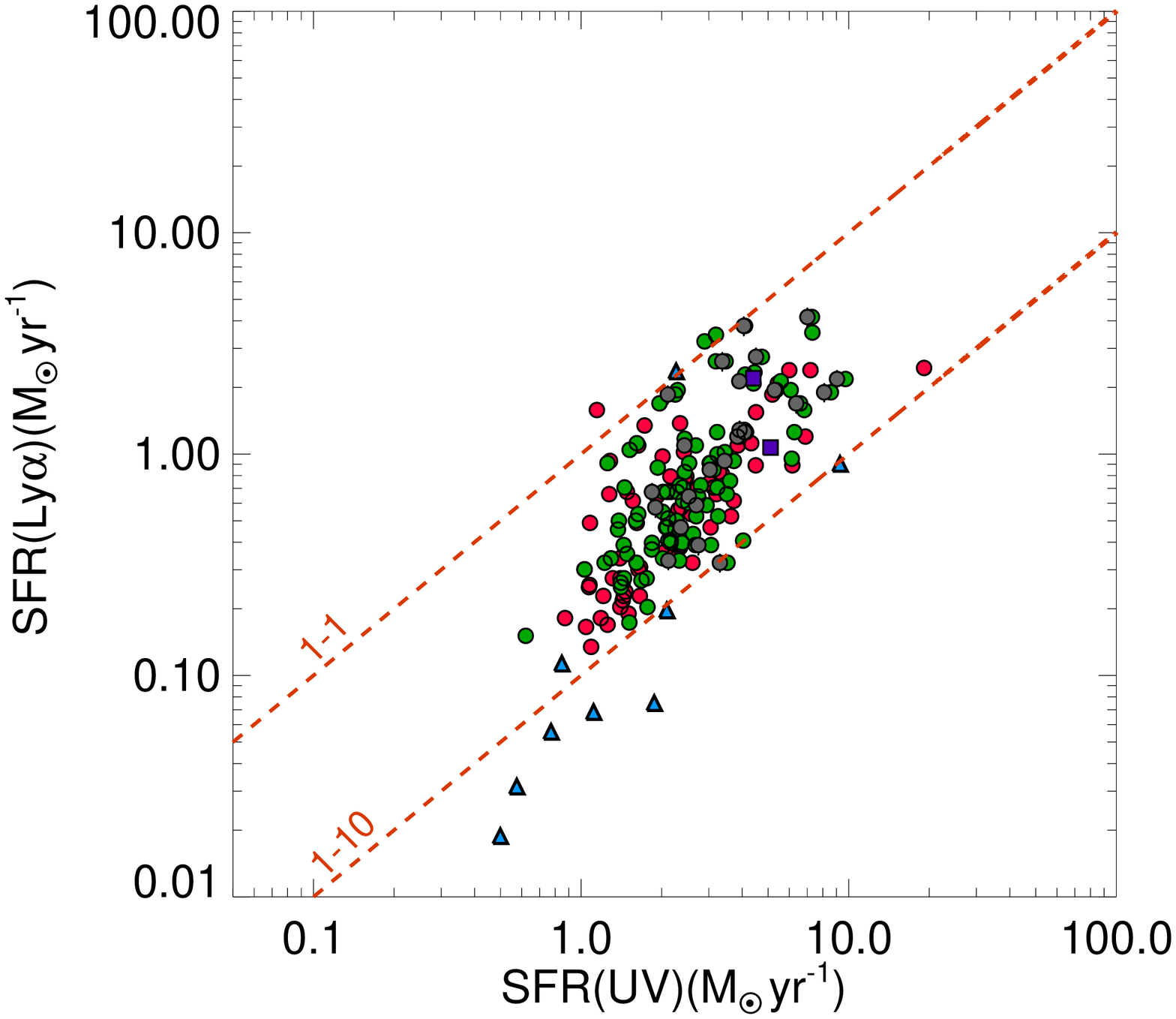}
   \caption{Comparison between SFR indicators. The SFR values are calculated using \citet{kennicutt98} calibration. The {\it left panel} shows SFR(\ha) versus SFR(UV) both corrected for dust extinction derived from the Balmer decrement. The dashed line denotes the line of equality, while the dotted lines have a factor of 2 deviation. The {\it right panel} presents the observed SFR(\lya) as a function of SFR(UV), with no correction for dust extinction. The two dashed lines marks the 1:1 and 1:10 ratios for SFR$_{Ly\alpha}$:SFR$_{UV}$. The symbols are the same as in Fig. \ref{fig:fesc_ebv}.}
   \label{fig:sfrlya_sfruv} 
\end{figure*}

Another possible source of the \lya\ excess is collisional excitation. If the electron temperature is high enough (typically higher than $\sim 2 \times 10^{4}$ K), collisions with thermal electrons lead to excitation and then radiative de-excitation. This process becomes non negligible compared to the recombination source in the case of high ISM temperatures. Furthermore, it has been suggested that a weak component of \lya\ emission can originate from ionization by the plasma of very-hot X-ray emitting filaments, as \citet{floranes12} observed a spatial correlation between the soft X-ray emission and the diffuse \lya\ component in the galaxy Haro 2. Given the small values of \fesc(\lya) that show a significant \lya\ enhancement, i.e.  \fesc(\lya) $< 5 \%$ for \fescnorm(\lya) $> 1$, one needs a relatively small contribution from the other emission processes to reproduce the deviation from expected attenuation by dust.

\section{\lya\ as a star formation rate indicator}
\label{sec:sfr}

In view of the important role that the \lya\ line will plays in the exploration of the distant universe with the upcoming new generation of telescopes such as the {\em JWST} and the {\em ELTs}, it is essential to know to what extent the observed \lya\ flux is representative of the intrinsic emission. The key capability for the detection of galaxies at $z > 7$ will be the rest-frame UV emission and, in the case of faint galaxies, the strongest emission line \lya. Typical high-redshift emission-line surveys rely on the strength of the \lya\ line, which is usually observed on top of a faint continuum. However, the interpretation of this line is hampered by many uncertainties related to complex transmission through the ISM and the IGM. 

In Figure \ref{fig:sfrlya_sfruv}, we first compare the SFR derived from \ha\ with the one based on the UV. Both quantities were corrected for dust extinction, were the stellar extinction affecting the UV continuum was assumed to be a factor of 2 lower than the gas phase extinction derived from the \ha/\hb\ ratio \citep{calzetti00}. The \ha\ emission is the result of the recombination of hydrogen atoms that have been previously ionized by the radiation of hot OB stars which have short lifetimes. It traces therefore the quasi instantaneous star formation rate on the scale of few Myrs. The UV (non ionizing) radiation is also emitted by less massive galaxies with longer life times -- typically few Gyrs -- and is indicative of the averaged SFR over the galaxy lifetime. We can see that the two SFR indicators are in broad agreement with a significant dispersion. Small differences are actually expected due to different star formation histories traced by these two emissions \citep[cf.][]{kennicutt98, schaerer03}

In order to assess the reliability of \lya\ as tracer of star formation, we now compare the SFR(\lya) with SFR(UV). The result is presented in the right panel of Fig. \ref{fig:sfrlya_sfruv}, which shows the observed SFRs derived using the \citet{kennicutt98} calibrations. The two dashed lines represent the ratios \sfruv/\sfrlya\ = 1 and 10. We see that the \lya\ indicator clearly underestimates the SFR when compared with the UV indicator. This is consistent with the discrepancy found in most of high-$z$ observations \citep[e.g.][]{yamada05,taniguchi05,gronwall07, tapken07,guaita10}. 
We note that the correction of SFR(\lya) with \fesc(\lya) gives a good estimate of the true SFR since it becomes equivalent to plotting SFR(\ha), given the definition of \fesc(\lya) (cf. Equation \ref{eq:fesc}). However, obtaining the \lya\ escape fraction for high-$z$ galaxies remains very challenging.

\begin{figure}[htbp]
   \centering
   \includegraphics[width=10.2cm]{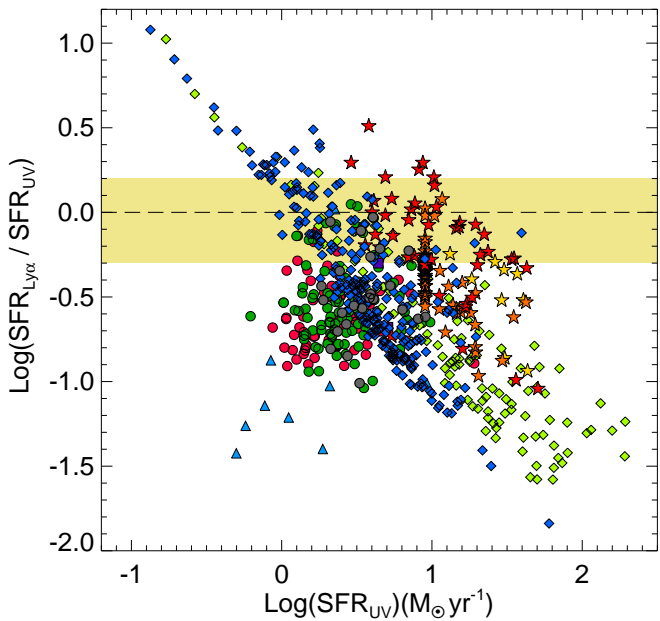}
   \caption{ SFR(\lya) to SFR(UV) ratio as a function of SFR(UV). The color-code for the local sample is the same as previous figures. Several high-z samples are also included: green and blue diamonds \citep{gronwall07,guaita10}; yellow, orange and red stars \citep{curtis12, taniguchi05,jiang13}. The SFRs are observed values, not corrected for dust extinction.}The dashed line corresponds to the SFR ratio of unity. The yellow region denotes the SFR ratio values that would be derived for different star formation histories (see text for details).
  \label{fig:sfr_ratio}
\end{figure}

We now compare the deviation of SFR(\lya) from the true SFR (which is assumed to be SFR(UV)) as a function of SFR(UV). The idea is to know when the observed \lya\ flux can be used as an SFR measurement. We plot in Figure \ref{fig:sfr_ratio} the logarithm of the observed ratio \sfrlya/\sfruv\ as a function of \sfruv. It is important here to compare our results to the high-z studies available in the literature. For this reason, we included in the plot of Fig. \ref{fig:sfr_ratio} additional samples described in the caption from $z \sim 2$ to $z \sim 6$. For reference, the dashed horizontal line represents the 1:1 ratio for  \sfruv/\sfrlya. The yellow region represents the range of values of \sfruv/\sfrlya\ that would be retrieved for different star formation histories, when the standard calibration from \citet{kennicutt98} is used (this calibration is based on the case of an evolved starburst already in its equilibrium phase). For a young burst or constant star formation that hasn't reached equilibrium yet (age $< 100$ Myr), SFR(\lya) is higher than SFR(UV), and a ratio of \sfrlya/\sfruv\ $ \sim 4$ can be observed \citep{schaerer03,verhamme08}, which is represented by the upper part of the yellow region. For older galaxies ($>$ 1 Gyr) with a constant star formation history, SFR(UV) becomes higher than SFR(\lya), which explains the lower part of the yellow region.

Clearly, the SFR measured from \lya\ is most of the time, and at all redshifts, lower than the true SFR. We can observe a general decline of the \sfrlya/\sfruv\ with increasing \sfruv. As discussed for Fig. \ref{fig:ewlya_sfruv} describing the decrease of \ewlya\ as a function of \sfruv, this can be interpreted as a decrease of \fesc(\lya) with increasing UV luminosity, mass, and possibly dust content, but also to the natural decline resulting from ploting 1/x versus x. This trend is predicted by the models, as \citet{garel12} find a lower \fesc(\lya) in higher-SFR galaxies at $z \sim 3-5$. An equivalent result is the observed decline of the fraction of LBGs that show \lya\ emission (defined as the \lya\ fraction $x_{\rm{Ly}\alpha}$) with increasing UV luminosity \citep[e.g.][]{stark10,schaerer11,schenker12}. Quantitatively, the studies find $x_{\rm{Ly}\alpha} \sim 50\%$ for $M_{UV}=-19$ and $x_{\rm{Ly}\alpha} \sim 10\%$ for $M_{UV}=-21$ at $3 < z <6$. 

More importantly, we note that this relationship in the different samples is shifted as the redshift increases, from $z \sim 0$ (triangles and circles) to $z$ \gsim 6 (stars) passing by $z \sim 2-3$ (diamonds). This can be seen as the result of an increasing \lya\ escape fraction as a function of redshift. Indeed, the ratio of \sfrlya\ over \sfruv\ represent a proxy for the \lya\ escape fraction, assuming constant star formation history. By comparing the observed \lya\ luminosity function with the intrinsic one at different redshifts, a clear evolution of \fesc(\lya) with redshift can be seen in \citet{hayes11}. The study relies on a compilation of \lya\ LFs extracted from the literature between $z \sim 0$ and 8, and integrated over homogeneous limits to obtain a \lya\ luminosity density, which is then compared to the intrinsic \lya\ luminosity density to obtain the sample-averaged volumetric \lya\ escape fraction. The intrinsic \lya\ luminosity density is obtained from dust-corrected \ha\ at $z \sim 2.3$ and from UV continuum at $z > 2.3$ where \ha\ observations are not available. The resulting \lya\ escape fraction increases from $\lesssim 1$\% at $z \sim 0$ to $\sim 40$\% at $z \sim 6$, comparable to the redshift range presented in Fig. \ref{fig:sfr_ratio}. Similarly, \citet{stark10} showed that the prevalence of \lya\ emitters amongst LBGs increases with redshift over $3 < z < 6$. This trend seems to continue down to $z \sim 0.3$, where \citet{cowie10} showed that only $5\%$ of their {\em GALEX}-NUV-selected galaxies are \lya\ emitters, while they represent $\sim 20\%$ of the $z \sim 3$ LBG sample of \citet{shapley03}.

This trend is mainly driven by the decrease of dust content of galaxies with redshift. The measured E(B-V) in the galaxy samples compiled in \citet{hayes11} evolves clearly with redshift. Furthermore, \citet{hayes11} combined the \fesc(\lya)-E(B-V) relation with their measured \lya\ and UV SFR densities, to predict the evolution of dust content with redshift. Independently of the measured E(B-V) they found a clear decrease of dust with increasing redshift up to $z \sim 6$. Such evolution has been demonstrated in other studies, by measuring the UV slope of LBGs at $2 < z < 7$ \citep[e.g.][]{hathi08,bouwens09}.

Considering the different techniques used to detect high-z galaxies, one should be aware of possible selection biases that could affect this kind of analysis. First, since the evolution of the SFR ratio is attributed to the evolution of \fesc(\lya), we should investigate whether this ratio is affected by other important factors. It has been shown for instance that the prevalence of high-EW sources increases with redshift \citep{ouchi08,atek11,shim11}, which could produce the observed redshift-evolution. Because, most of the high-$z$ LAEs are elected in narrow-band surveys, they could miss the low-EW objects that lie below the detection threshold, whereas they will be detected in lower-redshift studies that use spectroscopic selection for instance. However, the typical threshold used by narrow-band survey is $EW_{{\rm Ly}\alpha}$ = 20 \AA, and the distributions of $EW_{{\rm Ly}\alpha}$ at high-$z$ suggest that a maximum of 20\% of such galaxies can be missed. Another effect is that we are comparing galaxy samples with different UV luminosities while there is evidence that \lya\ is stronger in lower-luminosity galaxies as explained above \citep{ando06,verhamme08,pentericci09,stark10, schaerer11}. This is what we see in the same Fig. \ref{fig:sfr_ratio} where the, the SFR ratio is decreasing with increasing SFR(UV) for a given redshift slice. The result is consistent with both theoretical and observational evidences, where galaxies with higher UV luminosities are more obscured by dust and tend to have higher metallicities, masses, and larger reservoir of neutral gas \citep{reddy06,pirzkal07,gawiser07,overzier08,verhamme08, lai08}, all of which are contributing to lower the \lya\ escape fraction and therefore the SFR ratio. The region at the center of the plot where the different samples tend to overlap is most likely symptomatic of the general dispersion of \fesc(\lya) because of the multi-parameter nature of the resonant scattering process \citep{verhamme06, atek08, atek09b, hayes10b}. We here refer the reader to \citet{hayes11} for a detailed discussion about the other factors that can affect the \lya/UV ratio.

 As pointed out in Section \ref{sec:fesc}, the galaxy sample of the present study is clearly biased towards strong \lya\ emitters as opposed to high-z LBGs for instance. This favors galaxies that show high \lya\ escape fractions, which could explain the high number of galaxies with recombination ratios above the level expected from dust attenuation when compared to other selection methods \citep{kornei10, hayes11}. If anything, this selection bias should work against the anti-correlation observed in Fig. \ref{fig:sfr_ratio} between the SFR ratio and SFR(UV). Another selection effect is that at higher redshift, galaxies will have a higher UV luminosity, which make the increase of SFR ratio as a function of redshift not very obvious at a given UV (or SFR) luminosity. Ideally, we would assemble samples of galaxies at different redshift within the same UV luminosity bin. Nevertheless, as the redshift increases,  the SFR(UV) increases, but so does \fesc(\lya). For this reason we are able to clearly see mean \sfrlya/\sfruv\ getting closer to unity at high redshift.

\section{Conclusion}
\label{sec:conclusion}

We have analyzed a large sample \lya\ emitters at redshift $z \sim 0$ and $z \sim 0.3$. We combined our spectroscopic follow-up of $z \sim 0.3$ galaxies, originally detected by \citet{deharveng08} with {\it GALEX}, with various data obtained from the literature to investigate the influence of several physical parameters on the escape of \lya\ emission. After estimating the AGN contamination, we measured the oxygen abundance and the gas-phase extinction, using optical emission line ratios. We also derived the \lya\ escape fraction, \fesc(\lya), using the \ha\ flux and the nebular extinction. We summarize here the main conclusions we draw from the comparison of these parameters with the \lya\ and UV properties.
\begin{itemize}    
\item The \lya\ escape fraction or equivalent width does not show a correlation with metallicity. While one may expect metal-poor galaxies to show strong \lya\ emission as observed in \citet{cowie11}, our results show that this is not always the case, and are in agreement with the observations of local blue compact galaxies, such as \izw\ and SBS 0335. 
\smallskip
\item Looking at the \lya\ equivalent width as a function of extinction, we do not find the trend commonly observed in high-redshift samples \citep[e.g.][]{shapley03, pentericci09}. We explain the absence of correlation by the decoupling of the attenuation affecting the UV continuum and the multi-parameter process responsible for the attenuation the \lya\ line. In addition, high-redshift studies use a model-dependent extinction that refers to the stellar extinction, which can be very different from the gas-phase dust encountered by \lya\ photons.
\smallskip
\item The \lya\ escape fraction presents a clear correlation with the dust extinction, confirming earlier results \citep[e.g.][]{atek09b,kornei10,hayes11}, albeit with a large dispersion indicating that dust attenuation is not the only regulatory factor of \lya\ emission. A two-parameter fit of \fesc(\lya)-E(B-V) yields an extinction coefficient of $k_{Ly\alpha} \sim 6.67 \pm 0.52$ and an intercept point of C$_{Ly\alpha} = 0.22 \pm 0.03$. We see that even when no dust is present, the point of \fesc(\lya)=1 is only an upper limit and \lya\ emission can be attenuated by the resonant scattering into neutral gas. 
\smallskip
\item We also introduce the normalized \fesc(\lya), which corresponds to the deviation of the observed \fesc(\lya) from what is expected from the case of pure dust attenuation. At low extinction, \lya\ appears more attenuated than the UV continuum, most likely because of the scatting process. At higher extinction, \fesc(\lya) show above the predictions for standard dust attenuation. This can be the results of various mechanisms, including an enhancement of \lya\ due to the ISM geometry \citep{neufeld91, laursen12, verhamme12}, a different extinction law \citep{scarlata09}, a significant contribution from collisional excitation, or ionization by a hot plasma \citep{floranes12}.
\smallskip
\item In order to assess the reliability of the \lya\ emission line as a star formation indicator, we compared SFR(\lya) and SFR(UV), which has the advantage of being independent of the extinction. Overall, \lya\ tend to underestimate the SFR by a factor up to 10. We observe that the \sfrlya/\sfruv\ ratio is decreasing with increasing \sfruv, which can be interpreted as a decrease in \fesc(\lya) as a function of UV luminosity as observed in several studies. We also note that this trend is being shifted with increasing redshift because of the redshift-dependence of the \lya\ escape fraction.
\end{itemize}

\begin{acknowledgements}
We thank the anonymous referee for useful suggestions that improved the clarity of the paper and Len Cowie for providing us with his emission line measurements. HA and DK acknowledge support from the Centre National d'Etudes Spatiales (CNES). HA and JPK are supported by the European Research Council (ERC) advanced grant ``Light on the Dark'' (LIDA). This work is based on observations made with ESO Telescopes at La Silla Observatories under program ID 082.B-0392. 
\end{acknowledgements}

\bibliographystyle{aa}
\bibliography{galex}

\begin{figure*}[htbp]
   \centering
   \includegraphics[width=19cm]{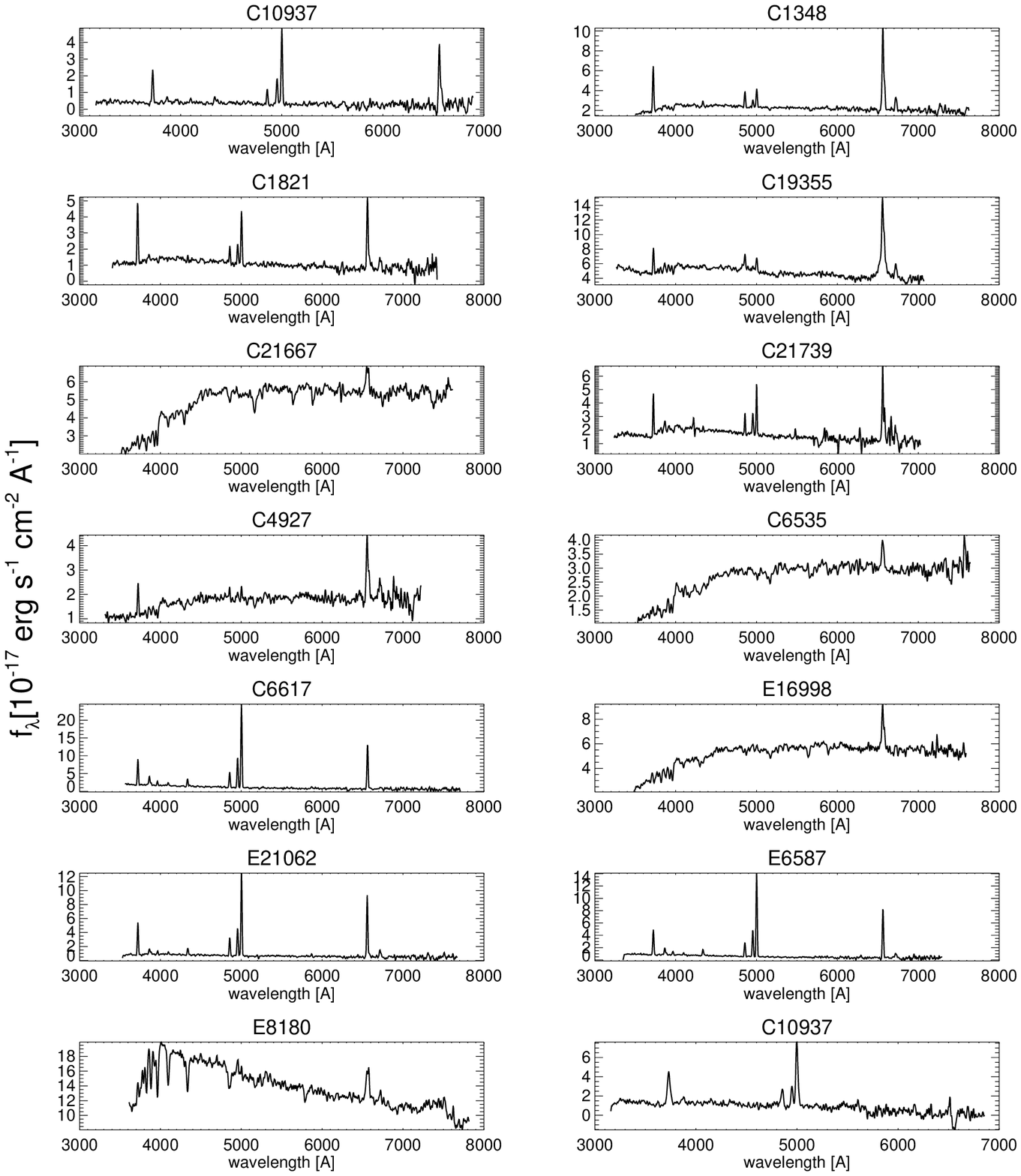}
   \vspace{-2cm}
   \caption{Same as Fig. \ref{fig:galex_spectra}}
   \label{fig:galex_spectra2}
\end{figure*}

\end{document}